\documentclass[12pt]{article} \textwidth 6.25in \textheight 9in
\usepackage{latexsym}
\usepackage{graphics}
\usepackage{graphicx}
\usepackage{latexsym}
\usepackage{amssymb,amsmath}
\usepackage{epsfig}
\usepackage{url}

\makeatletter

\newcommand{\singlespacing}{\let\CS=\@currsize\renewcommand{\baselinestretch}{1}\tiny\CS}
\newcommand{\oneandahalfspacing}{\let\CS=\@currsize\renewcommand{\baselinestretch}{1.25}\tiny\CS}
\newcommand{\doublespacing}{\let\CS=\@currsize\renewcommand{\baselinestretch}{1.5}\tiny\CS}
\usepackage[noadjust]{cite}
\newtheorem{theorem}{\textbf{Theorem}}
\newtheorem{definition}{\textbf{Definition}}
\newtheorem{lemma}{\textbf{Lemma}}
\newtheorem{property}{\textbf{Property}}
\newtheorem{corollary}{\textbf{Corollary}}
\usepackage{times}
\usepackage{epsf}
\usepackage{epsfig}
\usepackage{amsfonts}
\usepackage{amstext}
\usepackage{ifthen}
\usepackage{multirow}
\usepackage{subfigure}

\newcommand{\Expt}{\mbox{${\mathbb E}$} }
\renewcommand{\vec}[1]{\mbox{\boldmath$#1$}}

\begin{document}
\title{On Multistage Successive Refinement for Wyner-Ziv Source Coding with Degraded Side Informations}
\author{Chao Tian and Suhas Diggavi
\thanks{Chao Tian and Suhas Diggavi are with the School of Computer and Communication Sciences Swiss Federal Institute of Technology (EPFL).
 Email: {\tt \{chao.tian,suhas.diggavi\}@epfl.ch}}}

\maketitle
\begin{abstract}
We provide a complete characterization of the rate-distortion region
for the {\em multistage} successive refinement of the Wyner-Ziv source
coding problem with degraded side informations at the
decoder. Necessary and sufficient conditions for a source to be
successively refinable along a distortion vector are subsequently
derived. A source-channel separation theorem is provided when the descriptions are sent over independent channels for the multistage case. Furthermore, we introduce the notion of generalized
successive refinability with multiple degraded side
informations. This notion captures whether progressive encoding to
satisfy multiple distortion constraints for different side
informations is as good as encoding without progressive
requirement. Necessary and sufficient conditions for generalized
successive refinability are given. It is shown that the following two
sources are generalized successively refinable: (1) the Gaussian
source with degraded Gaussian side informations, (2) the doubly
symmetric binary source when the worse side information is a
constant. Thus for both cases, the failure of being successively
refinable is only due to the inherent uncertainty on which side
information will occur at the decoder, but not the progressive
encoding requirement.
\end{abstract}

\section{Introduction}
\label{sec:intro}

The notion of successive refinement of information was introduced by
Koshelev \cite{Koshelev:80} and by Equitz and Cover
\cite{EquitzCover:91}, whose interest was to determine whether the
requirement of encoding a source progressively necessitates a higher
rate than encoding without the progressive requirement.  A source is
said to be successively refinable if encoding in multiple stages
incurs no rate loss as compared with optimal rate-distortion encoding
at the separate distortion levels.  Rimoldi \cite{Rimoldi:94} later
provided a complete characterization of the rate-distortion region for
this problem.

In another seminal paper, Wyner and Ziv \cite{Wynerziv:76}
characterized the rate-distortion function for encoding a source when
the decoder alone has access to side information correlated with the
source. The notion of successive refinement was combined with the
presence of side information by Steinberg and Merhav
\cite{Steinbergmerhav:04}, who formulated the problem of successive
refinement with {\em degraded side informations} at the decoder.  The
degradedness roughly means that the decoder receiving the higher rate
bit-stream also has access to the ``better quality'' side
information. 
More formally, this means the source and
side-informations arranged in the descending order according to the
rate of bitstream form a Markov chain.  The notion of successive
refinability with degraded side informations was consequently defined,
which answers the question whether such a progressive encoding causes
rate loss as compared with a single stage Wyner-Ziv coding. In this
context, the main result in \cite{Steinbergmerhav:04} was the
characterization of the rate-distortion region and the necessary and
sufficient conditions for successive refinability for
{\em two-stage} systems. The characterization for more than two stages was
left open. An achievable region was indeed given, however, the
converse proof was not found\footnote{In fact, the complete
rate-distortion region for multi-stage system with {\em identical}
side information was given, however this only addresses a special case
in the framework.}.

In this work we extend these ideas in several ways. First, the
question left open by Steinberg and Merhav is resolved, which is the
characterization of the rate-distortion region for the successive
refinement under the Wyner-Ziv setting, for any finite number of
degraded side informations. This is accomplished by an alternative
representation of the rate region based on rate-sums.  This
characterization overcomes the difficulty perhaps encountered by
Steinberg and Merhav, in proving the converse for the general
multistage achievable region they found. The achievable region
provided in \cite{Steinbergmerhav:04} is then analyzed and shown to be
equivalent to the rate-distortion region. Necessary and sufficient
conditions for a source to be successively refinable are derived.

The notion of successive refinability introduced by Steinberg and
Merhav can be quite restrictive.  This can be understood in the
context of work of Heegard and Berger \cite{Heegardberger:85}, as well
as Kaspi \cite{Kaspi:94}, who studied the problem of source coding
when a correlated side information may or may not be available at the
decoder. In particular, it was shown that when transmission was to
multiple decoders with degraded side informations, the rate distortion
function could exceed the Wyner-Ziv rate needed for the decoder with
the ``stronger" side information, as well as that needed for the
decoder with the ``weaker" side information. As such, sources can fail
to be successively refinable (with side information) simply due to
this reason.  This motivates our definition of generalized successive
refinability of sources when decoders have access to multiple side
informations. In this notion we only require the sum-rate of the
progressive encoding to match the Heegard-Berger rate for degraded
side informations, instead of the Wyner-Ziv rate.  Necessary and
sufficient conditions for a source to have this property are then
given.  This notion of generalized successive refinability is applied
to Gaussian sources with jointly Gaussian side informations and
quadratic distortion measure. It is shown that the Gaussian source is
actually successively refinable in the generalized sense, though it
fails to be successively refinable in the strict sense as defined by
Steinberg and Merhav in most cases. An explicit calculation is also
given for the doubly symmetric binary source (DSBS) under Hamming
distortion measure, when the worse side information is a constant,
which we show is also successively refinable in the generalized
sense. The explicit calculation of the rate-distortion region for the
DSBS source in fact gives the Heegard-Berger rate-distortion function,
which was not found as of our knowledge despite several attempts
\cite{Heegardberger:85,Kerpez:87,FlemingEffros:03,Flemingthesis}.

The result can be generalized to the scenario when the descriptions are transmitted over $N$ independent discrete memoryless channel (DMC). In a more recent work \cite{SteinbergMerhav:06}, Steinberg and Merhav showed a source-channel separation result holds for the two-stage case. In light of the our new result, it can be shown that such separation holds for the multistage case as well. 
   
The rest of the paper is organized as follows. In Section
\ref{sec:prelim} we define the problem and establish the notation.  In
Section \ref{sec:MultiSucRef}, a characterization is provided for the
rate-distortion region with an arbitrary finite number of stages,
therefore the question left open in \cite{Steinbergmerhav:04} is
resolved. Section \ref{sec:GenSucRef} begins with the necessary and
sufficient conditions for a source to be successive refinable, then the
notion of generalized successive refinability is introduced and
investigated. The Gaussian example is explored in Section
\ref{sec:Gauss}, and the doubly symmetric binary source is
investigated in \ref{sec:DSBS}. Section \ref{sec:disc} concludes this
paper with a brief discussion. Proof details are given in the
appendices.

\section{Notation and Problem Statement}
\label{sec:prelim}

Let $\mathcal{X}$ be a finite set and let $\mathcal{X}^n$ be the set
of all $n$-vectors with components in $\mathcal{X}$.  Denote an
arbitrary member of $\mathcal{X}^n$ as $x^n=(x_1,x_2,...,x_n)$, or
alternatively as $\vec{x}$ when the dimension $n$ is clear from the
context. Upper case is used for random variables and vectors. A
discrete memoryless source (DMS) $(\mathcal{X},P_X)$ is an infinite
sequence $\{X_i\}_{i=1}^{\infty}$ of independent copies of a random variable
$X$ in $\mathcal{X}$ with a generic distribution $P_X$
\begin{eqnarray}
P_X(x^n)=\prod_{i=1}^nP_X(x_i).
\end{eqnarray}
Similarly, let
$(\mathcal{X},\mathcal{Y}_1,\mathcal{Y}_2,...,\mathcal{Y}_N,P_{XY_1Y_2,...,Y_N})$
be a discrete memoryless multisource with generic distribution
$P_{XY_1Y_2,...,Y_N}$, where $N$ is the number of coding stages.

\begin{figure}[tb]
  \centering
    \includegraphics[width=14cm]{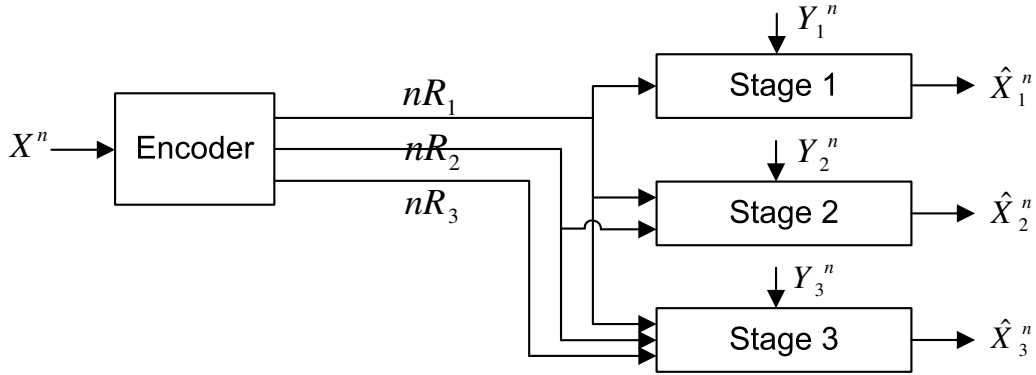}
\caption{A three-stage successive refinement system with side informations. \label{fig:sysdiag} The side informations are degraded in the sense that $X\leftrightarrow Y_3\leftrightarrow Y_2\leftrightarrow Y_1$.}
\end{figure}

Let $\hat{\mathcal{X}}$ be a finite reconstruction alphabet, and let 
\begin{equation}
\mathit{d}:\mathcal{X}\times\hat{\mathcal{X}}\rightarrow \left[0,\infty\right)
\end{equation}
be a distortion measure. For simplicity, we will assume the decoders
at all the stages use the same reconstruction alphabet and have the same distortion measure.  
The generalization to different distortion measures and reconstruction alphabets
is quite simple. The per-letter distortion of a vector is defined as
\begin{eqnarray}
\mathit{d}(\vec{x},\vec{\hat{x}})=\frac{1}{n}\sum_{i=1}^n
\mathit{d}(x_i,\hat{x}_i), \quad \forall \vec{x}\in \mathcal{X}^n,
\quad \vec{\hat{x}}\in \hat{\mathcal{X}}^n.
\end{eqnarray}
All the $\log$ function in this work is taken to be base 2.

\begin{definition}
An $(n,M_1,M_2,...,M_N,D_1,D_2,...,D_N)$ successive refinement (SR) code for source $X$ with
side information $(Y_1,Y_2,...,Y_N)$ consists of $N$ encoding
functions $\phi_m$, $m=1,2,...,N$, and $N$ decoding functions
$\psi_m$, $m=1,2,...,N$:
\begin{eqnarray}
\phi_m&:&\mathcal{X}^n\rightarrow I_{M_m}\\
\psi_m&:&I_{M_1}\times I_{M_2} \times ...\times I_{M_m} \times \mathcal{Y}_m^n \rightarrow \hat{\mathcal{X}^n},
\end{eqnarray}
where $I_k=\{1,2,...,k\}$, such that 
\begin{eqnarray}
\Expt \mathit{d}(X^n,\psi_m(\phi_1(X^n),\phi_1(X^n),...,\phi_m(X^n),Y_m^n))\leq D_m,
\end{eqnarray}
where $\Expt$ is the expectation operation. 
\end{definition}

\begin{definition}
A rate vector $\vec{R}=(R_1,R_2,...,R_N)$ is said to be
$\vec{D}=(D_1,D_2,...,D_N)$ achievable, if for every $\epsilon>0$
there exists for sufficient large $n$ an
$(n,M_1,M_2,...,M_N,D_1+\epsilon,D_2+\epsilon,...,D_N+\epsilon)$ code
with
\begin{equation}
R_m+\epsilon \leq \frac{1}{n}\log M_m, \quad m=1,2,...,N.
\end{equation}
\end{definition}

A three-stage example is given in Fig. \ref{fig:sysdiag}.
Denote the collection of all the $\vec{D}$ achievable rate vectors as $\mathcal{R}(\vec{D})$, and this is the region to be
characterized. When the side informations have arbitrary dependence
among them, the problem appears to be difficult. As in
\cite{Steinbergmerhav:04}, we consider only the case with a particularly
ordered degraded side informations, which is given by the Markov condition
$X\leftrightarrow Y_N \leftrightarrow Y_{N-1} \leftrightarrow
...\leftrightarrow Y_1$. One of our main results is the complete
characterization of this region, given in the next section.

\begin{figure}[tb]
  \centering
    \includegraphics[width=14cm]{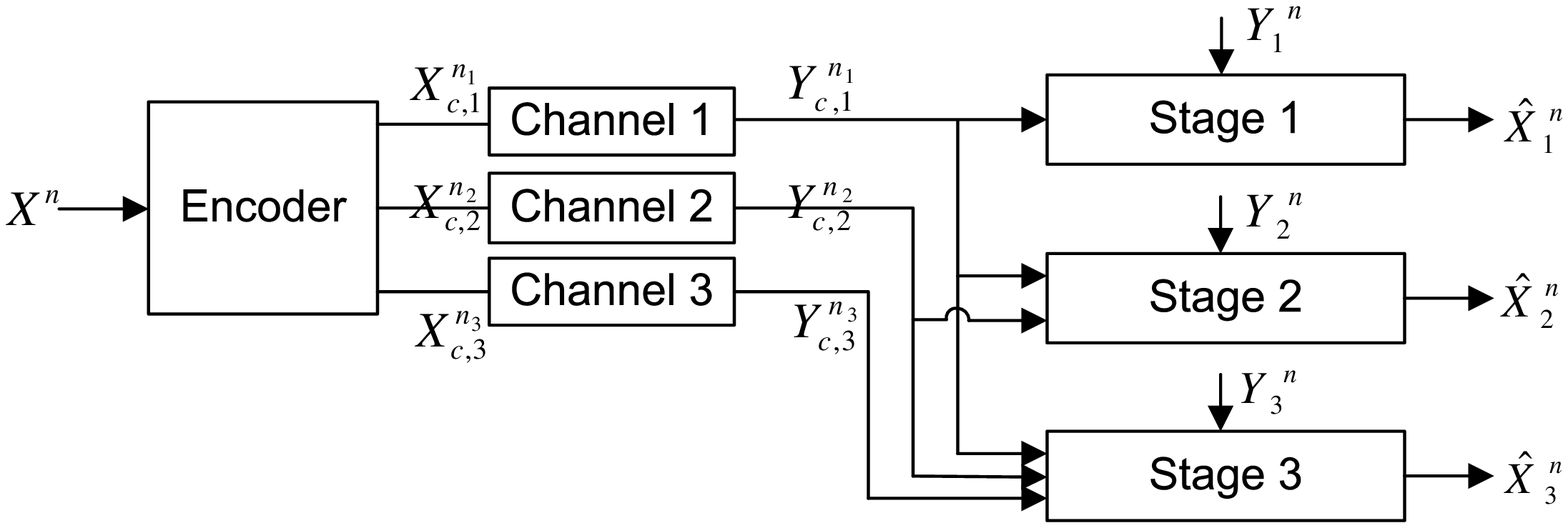}
\caption{The corresponding source channel coding problem for the source coding system depicted in Fig. \ref{fig:sysdiag}. \label{fig:SCsysdiag}.}
\end{figure}

We can further consider the case when the descriptions are transmitted over $N$ independent discrete memoryless channel (DMC) (see Fig \ref{fig:SCsysdiag}). For simplicity, instead of using the more general model where the channels are cost-constrained as in \cite{SteinbergMerhav:06}, we only consider channels without constraints; however, such an extension can be done without much difficulty.
\begin{definition}
An $(n,n_1,n_2,...,n_N,D_1,D_2,...,D_N)$ source-channel SR (SC-SR) code for source $X$ with
side information $(Y_1,Y_2,...,Y_N)$ for independent channels given by $P_{Y_{c,m}|X_{c,m}}$, $m=1,2,...,N$, consists of $N$ encoding
functions $\phi_m$, $m=1,2,...,N$, and $N$ decoding functions
$\psi_m$, $m=1,2,...,N$:
\begin{eqnarray}
\phi_m&:&\mathcal{X}^n\rightarrow \mathcal{X}^{n_m}_{c,m}\\
\psi_m&:&\mathcal{Y}^{n_1}_{c,1}\times \mathcal{Y}^{n_2}_{c,2} \times ...\times \mathcal{Y}^{n_m}_{c,m} \times \mathcal{Y}_m^n \rightarrow \hat{\mathcal{X}^n},
\end{eqnarray}
such that 
\begin{eqnarray}
\Expt \mathit{d}(X^n,\psi_m(\vec{Y_{c,1}},\vec{Y_{c,2}},...,\vec{Y_{c,3}},\vec{Y_m}))\leq D_m.
\end{eqnarray}
\end{definition}

\begin{definition}
A distortion vector $\vec{D}=(D_1,D_2,...,D_N)$ is said to be SC-SR achievable for source $P_{XY_1Y_2,...,Y_N}$ and channels $P_{Y_{c,m}|X_{c,m}}$, $m=1,2,...,N$, under bandwidth expansion factor $(\rho_1, \rho_2,...,\rho_N)$, if for every $\epsilon>0$ there exists for sufficient large $n$ an
$(n,n\rho_1,n\rho_2,...,n\rho_N,D_1+\epsilon,D_2+\epsilon,...,D_N+\epsilon)$ SC-SR code. The  achievable SC-SR distortion region $\mathcal{D}(\rho_1,\rho_2,...,\rho_N)$ is the collection of all the SC-SR achievable distortion vectors under the given bandwidth expansion factors.
\end{definition}

\section{The Characterization of the Rate-distortion Region with Degraded Side Information}
\label{sec:MultiSucRef}

Define the region $\mathcal{R}^*(\vec{D})$ to be the set of all rate
vectors $\vec{R}=(R_1,R_2,...,R_N)$ for which there exists $N$ random
variables $(W_1,W_2,...,W_N)$ in finite alphabets
$\mathcal{W}_1,\mathcal{W}_2,...,\mathcal{W}_N$ such that the
following condition are satisfied.

\begin{enumerate}\label{enm:conditions}
\item $(W_1,W_2,...,W_N)\leftrightarrow X \leftrightarrow Y_N
\leftrightarrow Y_{N-1} \leftrightarrow... \leftrightarrow Y_1$.
\item There exist deterministic maps $f_m: \mathcal{W}_m\times \mathcal{Y}_m \rightarrow \hat{\mathcal{X}}$ 
such that
\begin{eqnarray}
\Expt\mathit{d}(X,f_m(W_m,Y_m))\leq D_m, \quad 1\leq m \leq N.
\end{eqnarray}
\item The alphabet sizes satisfies 
\begin{eqnarray}
|\mathcal{W}_1|&\leq& |\mathcal{X}|+2N-1\nonumber\\
|\mathcal{W}_m|&\leq& |\mathcal{X}|\prod_{i=1}^{m-1}|\mathcal{W}_i|+2N-2m-1, \quad m=2,3,...,N.
\end{eqnarray}
\item The non-negative rate vectors satisfies:
\begin{eqnarray}
\label{eqn:rates}
\sum_{i=1}^m R_i\geq \sum_{i=1}^m I(X;W_m|W_1,W_2,...,W_{m-1},Y_m), \quad 1\leq m \leq N. \label{eqn:ratevectcond}
\end{eqnarray} 
\end{enumerate}
where we have used the convention that $W_0=\emptyset$, {\em i.e.,} the null set.

\paragraph*{Remark 1} Because of the conditioning on $W_1,W_2,...,W_{m-1}$ in the
rate expressions, it is clear that the function $f_m(W_m,Y_m)$ can
also be written as $f_m'(W_1,W_2,...,W_m,Y_m)$ without essential
difference on the definition of the region. This equivalence will be
used in the explicit calculation of the rate-distortion region in
Section \ref{sec:Gauss} and \ref{sec:DSBS}. Furthermore, more structure can be built into the random variables $W_m$'s, such that the Markov chain holds as follows $W_1\leftrightarrow W_2 \leftrightarrow ...\leftrightarrow W_N\leftrightarrow X \leftrightarrow Y_N
\leftrightarrow Y_{N-1} \leftrightarrow... \leftrightarrow Y_1$; however, such additional structure requires an increase in the cardinality of the alphabets (see discussions in \cite{Steinbergmerhav:04}).

The following theorem establishes the rate-distortion region, 
which is one of the main results of the paper.
\begin{theorem}
\label{theorem:region}
For any discrete memoryless stochastically degraded source $X
\leftrightarrow Y_N \leftrightarrow Y_{N-1}
\leftrightarrow... \leftrightarrow Y_1$
\begin{eqnarray}
\mathcal{R}(\vec{D})=\mathcal{R}^*(\vec{D}).
\end{eqnarray}
\end{theorem}

The achievability of the region is quite straightforward.  The $m$-th
 stage codebook of overall size
 $2^{n(I(X;W_m|W_1,W_2,...,W_{m-1})+\epsilon_m)}$ is generated
 uniform-randomly from
 $T_{[W_m|\vec{w_1},\vec{w_2},...,\vec{w_{m-1}}]\delta}^n$, where
 $T_{[W_m|\vec{w_1},\vec{w_2},...,\vec{w_{m-1}}]\delta}^n$ denotes the
 set of $\delta$-typical sequences given lower-hierarchy codewords
 $(\vec{w_1},\vec{w_2},...,\vec{w_{m-1}})$.  These codewords are then
placed into  $2^{n(I(X;W_m|W_1,W_2,...,W_{m-1},Y_m)+2\epsilon_m)}$ bins
using a  uniform distribution.  The
 decoder block-decodes $\vec{W_m}$ in the $m$-th stage (using the side
 information), which is conditional on the lower hierarchy codewords;
 since the side informations are degraded, each higher hierarchy can
 always decode the lower-hierarchy codewords.  From the above
 interpretation, it is seen that the proof of the achievability of the
 region essentially uses the hierarchy of random codes as in the proof
 of the two stage case in \cite{Steinbergmerhav:04}. Thus we will
 focus on the converse part of the proof of the theorem, which is
 given in Appendix \ref{appsec:AppI}.

A source-channel separation result is now stated, and the proof is given in Appendix \ref{append:SCcoding}. 
\begin{theorem}
\label{theorem:SCregion}
For any discrete memoryless stochastically degraded source $X
\leftrightarrow Y_N \leftrightarrow Y_{N-1}
\leftrightarrow... \leftrightarrow Y_1$, and N independent discrete memoryless channels given by $P_{Y_{c,m}|X_{c,m}}$, $m=1,2,...,N$, the distortion vector $\vec{D}=(D_1,D_2,...,D_n)$ is achievable under bandwidth expansion factors $(\rho_1,\rho_2,...,\rho_N)$, if and only if there exist random variables $(W_1,W_2,...,W_N)$ in finite alphabets  $\mathcal{W}_1,\mathcal{W}_2,...,\mathcal{W}_N$ satisfying conditions 1), 2), 3) in the definition of $\mathcal{R}^*(\vec{D})$ and furthermore, 
\begin{eqnarray}
\label{eqn:SCrates}
\sum_{i=1}^m \rho_iC_i\geq \sum_{i=1}^m I(X;W_m|W_1,W_2,...,W_{m-1},Y_m), \quad 1\leq m \leq N, \label{eqn:SCratevectcond}
\end{eqnarray}
where $C_i$ is the channel capacity of channel $i$.
\end{theorem}

The rate region given in Theorem \ref{theorem:region} is in a different form than the achievable
region given in \cite{Steinbergmerhav:04}. Here
$\mathcal{R}^*(\vec{D})$ is given in terms of the sum-rate at each
stage, including rates at the previous stages, the sufficiency of
which was formally established in \cite{Effros:99}. The achievable
region in \cite{Steinbergmerhav:04}, denoted as
$\hat{\mathcal{R}}^*(\vec{D})$ here, involves $(N+1)N/2$ random
variables, and is given in terms of individual rate $R_m$ at each
stage. It is provided below for ease of comparison:
$\hat{\mathcal{R}}^*(\vec{D})$ is defined as the set of all rate
vectors $(R_1,R_2,...,R_N)$ for which there exists a collection of
$(N+1)N/2$ random variables $\{V_{i,j},1\leq i \leq N, i\leq j \leq N
\}$, where $V_{i,j}$ is taking values in a finite set
$\mathcal{V}_{i,j}$, such that the following conditions are satisfied.

\begin{enumerate}
\item $\{V_{i,j},1\leq i \leq N, i\leq j \leq N \}\leftrightarrow X
\leftrightarrow Y_N \leftrightarrow Y_{N-1}
\leftrightarrow... \leftrightarrow Y_1$.
\item There exist deterministic maps $
f_m: \mathcal{V}_{m,m}\times \mathcal{Y}_m \rightarrow \hat{\mathcal{X}}
$ 
such that
\begin{eqnarray}
\Expt\mathit{d}(X,f_m(V_{m,m},Y_m))\leq D_m,\quad 1\leq m \leq N.
\end{eqnarray}
\item The rate vectors satisfies:
\begin{align}
R_1&\geq I(X;V_{1,1}|Y_1)+\sum_{k=2}^NI(X;V_{1,k}|V_{1,1},V_{1,2},...,V_{1,k-1},Y_k)\\
R_m&\geq I(X;V_{m,m}|\{V_{i,j},\ 1\leq i <m,\ i \leq j \leq m\},Y_m)\nonumber\\
&\qquad+\sum_{k=m+1}^NI(X;V_{m,k}|\{V_{i,j},1\leq i \leq m,\ i \leq j \leq k-1\},Y_k),\quad 2\leq m \leq N.
\end{align} 
\end{enumerate}

It is clear that the characterization $\mathcal{R}^*(\vec{D})$ given in Theorem \ref{theorem:region} is more concise. However, it can indeed be shown that these two regions are equivalent, and we establish this equivalence as a theorem.
\begin{theorem}
For any discrete memoryless stochastically degraded source $X
\leftrightarrow Y_N \leftrightarrow Y_{N-1}
\leftrightarrow... \leftrightarrow Y_1$
\begin{eqnarray}
\hat{\mathcal{R}}^*(\vec{D})=\mathcal{R}^*(\vec{D})=\mathcal{R}(\vec{D}).
\end{eqnarray}
\end{theorem}

\begin{figure}[tb]
  \centering
    \includegraphics[width=5cm]{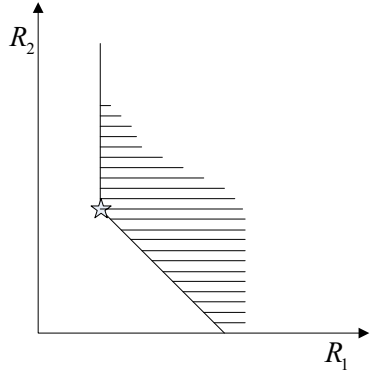}
\caption{An example when the achievability of the two regions are
equivalent, but the two regions are not the same. One region is
singleton point labeled using the star, the other region is the shaded
region including this singleton point. \label{fig:region}}
\end{figure}

The second equality obviously follows from Theorem 1.  Theorem 3 is
proved in Appendix \ref{appsec:AppII}, which might be of interest for the following
reason. In \cite{Steinbergmerhav:04}, a proof for a similar but
different claim was given for the special case of $N=2$, which showed
that the {\em achievability} of $\hat{\mathcal{R}}^*(\vec{D})$ and
$\mathcal{R}^*(\vec{D})$ are equivalent. However, this does not
directly imply that the two regions are equivalent; see
Fig. \ref{fig:region} for such an example. In our proof, the fact that
$\mathcal{R}^*(\vec{D})= \mathcal{R}(\vec{D})$ is used; and since
$\hat{\mathcal{R}}^*(\vec{D})$ is an achievable region, we have
trivially $\mathcal{R}^*(\vec{D})\supseteq
\hat{\mathcal{R}}^*(\vec{D})$. However, without invoking
$\mathcal{R}^*(\vec{D})= \mathcal{R}(\vec{D})$, it appears difficult
to prove this inclusion.  Interestingly, for $N=2$, it is indeed
possible to prove Theorem 2 without invoking
$\mathcal{R}^*(\vec{D})=\mathcal{R}(\vec{D})$, and this alternative
proof is also included in Appendix \ref{appsec:AppII}.

The following observation might shed some light on why a direct proof
of $\hat{\mathcal{R}}^*(\vec{D})=\mathcal{R}(\vec{D})$ might be
difficult, and it also provides the necessary intuition in proving
Theorem 3. Consider the case $N=3$, the random variable $V_{1,3}$ is
the information that the first stage encoded for the third
stage. However, if the second stage still has to encode $V_{2,2}$ with
a nonzero rate, then the encoder can not encode $V_{2,2}$ conditioned
on $V_{1,3}$, since the second stage decoder will not be able to
decode $V_{1,3}$. Furthermore $V_{1,3}$ does not help in the
second stage decoder either. As such the encoder might as well encode $V_{1,3}$ after
$V_{2,2}$ is encoded, which can then be conditioned on $V_{2,2}$ to
reduce the rate. Thus the optimal scheme is to encode the first stage
random variable $V_{1,1}$; if there is additional bit budget left in
the first stage, then adjust and encode $V_{1,2}$ conditioned on
$V_{1,1}$ until $V_{1,2}=V_{2,2}$; and if there is still additional
bit budget left, then adjust and encode $V_{1,3}$ conditioned on
$(V_{1,1},V_{2,2})$ until $V_{1,3}=V_{3,3}$, etc.; this process
carries for each stage sequentially. Thus the majority of the
$N(N+1)/2$ random variables are in fact null random variables, which
reflect the change of the coding strategy at boundary points. This
inherent change of encoding strategy appears to pose difficulty in
proving the converse using $\hat{\mathcal{R}}^*(\vec{D})$.

The example in Fig. \ref{fig:region} can also be explained by
introducing the following useful property.
\begin{property}
A region $\mathcal{K}$ is said to be {\em sum-incremental}, if the
following is true: if $\vec{R}\in \mathcal{K}$, then for any
non-negative rate vector $\vec{R'}$ that satisfies
$\sum_{i=1}^mR_i'\geq \sum_{i=1}^mR_i$ for all $1\leq m \leq N$,
$\vec{R'}\in \mathcal{K}$.
\end{property}
 
It was shown in \cite{Effros:99} that for successive refinement coding
without side information, the rate region is sum-incremental. Using
the same method, it can be shown that it is also true for the
rate-distortion region $\mathcal{R}(\vec{D})$ of successive refinement
coding in the Wyner-Ziv setting. Intuitively, this property states
that ``it does not matter how you divide up the rate between layers of
the (successively refining) descriptions, as long as the sum-rate of
first $m$ layers is sufficiently high for each $m=1,2,...,N$"
\cite{Effros:99}: we can simply move the rate in higher stages into
lower stages to form new codes.  The shaded region in
Fig. \ref{fig:region} is sum-incremental, well the singleton point
labeled by the star is not. Thus the shaded region can be a valid
rate-distortion region for the successive refinement problem, while
the singleton point is not, though the two regions imply the same
achievability result.  Now notice that it is quite difficult to prove
(even if not impossible) $\hat{\mathcal{R}}^*(\vec{D})$ is
sum-incremental, which suggests it will be difficult to prove
$\hat{\mathcal{R}}^*(\vec{D})=\mathcal{R}(\vec{D})$ directly.

\section{Strictly and Generalized Successive Refinability}
\label{sec:GenSucRef}

Extending the definition of successive refinability 
given in \cite{Steinbergmerhav:04} to an $N$-stage system, means the
following.

\begin{definition}
A source $X$ is said to be {\em $N$-step successively refinable} along
the distortion vector $\vec{D}=(D_1,D_2,...,D_N)$, with side
informations $(Y_1,Y_2,...,Y_N)$ if
\begin{eqnarray}
(R^*_{X|Y_1}(D_1),R^*_{X|Y_2}(D_2)-R^*_{X|Y_1}(D_1),...,R^*_{X|Y_N}(D_N)-R^*_{X|Y_{N-1}}(D_{N-1}))\in
\mathcal{R}(\vec{D})\label{eqn:strrefinable}
\end{eqnarray}
where $R^*_{X|Y}(\cdot)$ denotes the Wyner-Ziv rate distortion
function for source $X$ with side information $Y$ at the decoder.
\end{definition}

This definition of successive refinability will be referred to as {\em
strictly successive refinability}, for reasons that will become clear
shortly. The following theorem provides the conditions for $N$-stage
strictly successive refinability.

\begin{theorem}
\label{theorem:strrefinable}
A discrete memoryless stochastically degraded source $X
\leftrightarrow Y_N \leftrightarrow Y_{N-1}
\leftrightarrow... \leftrightarrow Y_1$ is $N$-step strictly
successively refinable along distortion vector $(D_1,D_2,...,D_N)$, if
and only if there exist random variables $(W_1,W_2,...,W_N)$ and
deterministic functions $f_m: \mathcal{W}_m\times \mathcal{Y}_m
\rightarrow \hat{\mathcal{X}}$ for $m=1,2...,N$ such that the
following conditions hold:
\begin{enumerate}
\item $R^*_{X|Y_m}(D_m)=I(X;W_m|Y_m)$ and 
$\Expt\mathit{d}(X,f_m(W_m,Y_m))\leq D_m$, $1\leq m \leq N$;
\item $(W_1,W_2,...,W_N)\leftrightarrow X \leftrightarrow Y_N
\leftrightarrow Y_{N-1}\leftrightarrow ... \leftrightarrow Y_1$;
\item $(W_1,W_2,...,W_{m-1})\leftrightarrow (W_m,Y_m)\leftrightarrow X$, $2\leq m\leq N$;
\item $I(W_i;Y_{m}|W_1,W_2,...,W_{i-1},Y_i)=0$, $1\leq i \leq m-1$, $2\leq m\leq N$.
\end{enumerate}
\end{theorem}

The conditions reduce to the corresponding conditions
for the two stage cases in \cite{Steinbergmerhav:04}.
Note that there are in fact a total of $N(N-1)/2$ equalities specified by condition
4). 

\noindent{\em Proof of Theorem \ref{theorem:strrefinable}}

For the necessity, assume (\ref{eqn:strrefinable}) holds. By Theorem
1, there exists random variables $(W_1,W_2,...,W_N)$ and maps $f_m:
\mathcal{W}_m\times \mathcal{Y}_m \rightarrow \hat{\mathcal{X}}$, such
that $(W_1,W_2,...,W_N)\leftrightarrow X \leftrightarrow Y_N
\leftrightarrow Y_{N-1}\leftrightarrow ... \leftrightarrow Y_1$, and
since (\ref{eqn:strrefinable}) holds, due to (\ref{eqn:ratevectcond}) we have,
\begin{eqnarray}
\label{eqn:k1}
R^*_{X|Y_m}(D_m)\geq \sum_{i=1}^m I(X;W_i|W_1,W_2,...,W_{i-1},Y_i), \quad 1\leq m \leq N,
\end{eqnarray}
and $\Expt\mathit{d}(X,f_m(W_m,Y_m))\leq D_m$, $1\leq m \leq N$.  From
(\ref{eqn:k1}), it follows that
\begin{small}
\begin{eqnarray}
&&R^*_{X|Y_m}(D_m)\geq \sum_{i=1}^m I(X;W_i|W_1,W_2,...,W_{i-1},Y_i)\nonumber\\
&\stackrel{(a)}{=}&[I(X;W_m|W_1,W_2,...,W_{m-1},Y_m)+\sum_{i=1}^{m-1} I(X;W_i|W_1,W_2,...,W_{i-1},Y_i)]\nonumber\\
&&+[\sum_{i=1}^{m-1}I(X;W_i|W_1,W_2,...,W_{i-1},Y_m)-\sum_{i=1}^{m-1}I(X;W_i|W_1,W_2,...,W_{i-1},Y_m)]\nonumber\\
&\stackrel{(b)}{=}&I(X;W_1,W_2,...,W_m|Y_m)+\sum_{i=1}^{m-1}[H(W_i|W_1,W_2,...,W_{i-1},Y_i)-H(W_i|W_1,W_2,...,W_{i-1},Y_i,X)\nonumber\\&&
-H(W_i|W_1,W_2,...,W_{i-1},Y_m)+H(W_i|W_1,W_2,...,W_{i-1},Y_m,X)]\nonumber\\
&\stackrel{(c)}{=}&I(X;W_1,W_2,...,W_m|Y_m)+\sum_{i=1}^{m-1}[H(W_i|W_1,W_2,...,W_{i-1},Y_i)-H(W_i|W_1,W_2,...,W_{i-1},Y_m)]\label{eqn:srcondition}\\
&\stackrel{(d)}{=}&I(X;W_1,W_2,...,W_m|Y_m)+\sum_{i=1}^{m-1}I(W_i;Y_m|W_1,W_2,...,W_{i-1},Y_i)\nonumber\\
&=&I(X;W_m|Y_m)+I(X;W_1,W_2,...,W_{m-1}|Y_m,W_m)+\sum_{i=1}^{m-1}I(W_i;Y_m|W_1,W_2,...,W_{i-1},Y_i)\nonumber\\
&\geq&R^*_{X|Y_m}(D_m)+\sum_{i=1}^{m-1}I(W_i;Y_m|W_1,W_2,...,W_{i-1},Y_i)\label{eqn:k3}\\
&\geq&R^*_{X|Y_m}(D_m)\label{eqn:k4}
\end{eqnarray}
\end{small}
where $(a)$ is by chain rule and adding and subtracting the same term, (b) follows by combining the first and third terms, $(c)$ is due to the Markov
chain relationship $(W_1,W_2,...,W_N)\leftrightarrow X \leftrightarrow Y_N
\leftrightarrow Y_{N-1} \leftrightarrow... \leftrightarrow Y_1$; $(d)$ is also due to the same Markov chain relationship which implies we can further condition
the last term in $(\ref{eqn:srcondition})$ with $Y_i$.
Next, inequality (\ref{eqn:k3}) is due to the fact that $(W_m,Y_m)$ is
sufficient to decode to a distortion $D_m$ while at the same time
satisfying the Markov condition $W_m\leftrightarrow X \leftrightarrow
Y_m$.  Because the beginning and the end of this chain of inequalities
are equal, all the inequalities must be equalities.  For
(\ref{eqn:k3}), the following two conditions must be true
\begin{eqnarray}
I(X;W_m|Y_m)=R^*_{X|Y_m}(D_m), \quad I(X;W_1,W_2,...,W_{m-1}|Y_m,W_m)=0
\end{eqnarray}
which implies $(W_1,W_2,...,W_{m-1})\leftrightarrow (W_m,Y_m) \leftrightarrow X$ for $2\leq m\leq N$.
For (\ref{eqn:k4}), it must be true that for $2\leq m \leq N$
\begin{eqnarray}
I(W_i;Y_m|W_1,W_2,...,W_{i-1},Y_i)=0, \quad 1\leq i \leq m-1.
\end{eqnarray}
This establishes the necessity. The sufficiency is of course trivial. The proof is completed.
\hfill$\square$

\paragraph*{Remark 2}: Following Remark 1 made after the definition of
$\mathcal{R}^*(\vec{D})$, we note that if the function $f_m(W_m,Y_m)$
is indeed given instead as $f_m'(W_1,W_2,...,W_m,Y_m)$, then the third
condition in Theorem \ref{theorem:strrefinable} will not appear in
this set of conditions, and the first condition should be modified as:
$R^*_{X|Y_m}(D_m)=I(X;W_1,W_2,...,W_m|Y_m)$ and
$\Expt\mathit{d}(X,f_m'(W_1,W_2,...,W_m,Y_m))\leq D_m$, $1\leq m \leq
N$.

In order to introduce the notion of generalized successive
refinability, we note that the problem considered in
\cite{Heegardberger:85},\cite{Kaspi:94} can be understood in the
framework being treated as the projection of rate vector
$\mathcal{R}(\vec{D})$ on the sum-rate $\sum_{i=1}^NR_i$ and ignoring
the individual rate; i.e., it is a relaxed version of the current
problem. Let us denote the sum-rate-distortion function to satisfy
distortion constraint vector $(D_1,D_2,...,D_m)$ with degraded side
information $(Y_1,Y_2,...,Y_m)$ as $R_{HB}(D_1,D_2,...,D_m)$, which
was given in \cite{Heegardberger:85}. Since $R_{HB}(D_1,D_2,...,D_m)$
degenerates to $R^*_{X|Y_m}(D_m)$ when all the other distortion
constraints $(D_1,D_2,...,D_{m-1})$ are set to be infinite, it is seen
that $R_{HB}(D_1,D_2,...,D_m)\geq R^*_{X|Y_m}(D_m)$. Because
$R_{HB}(D_1,D_2,...,D_m)$ is a lower bound for the sum-rate of
$\sum_{i=1}^mR_i$, if $R_{HB}(D_1,D_2,...,D_m)>R^*_{X|Y_m}(D_m)$ for any $m\in I_N$, then
the source is trivially not strictly successively refinable.

From the above discussion, it is seen that for a source to be strictly
successively refinable, two conditions are necessary. The first is
that $R_{HB}(D_1,D_2,...,D_m)=R^*_{X|Y_m}(D_m)$; and the second is
that in achieving $(D_1,D_2,...,D_m)$ for side information
$(Y_1,Y_2,...,Y_m)$, the encoding can be performed progressively
without rate loss. The first condition in fact provides a simple necessary condition
to check whether a source is successive refinable without directly
testing the conditions in Theorem \ref{theorem:strrefinable}, which
can be quite difficult because of the involvement of random variables
$W_i$.
\begin{theorem}
A necessary condition for a discrete memoryless stochastically
degraded source $X \leftrightarrow Y_N \leftrightarrow Y_{N-1}
\leftrightarrow... \leftrightarrow Y_1$ to be $N$-step strictly
successively refinable along distortion vector $(D_1,D_2,...,D_N)$, is
that $R_{HB}(D_1,D_2,...,D_m)=R^*_{X|Y_m}(D_m)$ for each $1\leq m\leq
N$.
\end{theorem}

This condition is in fact extremely strict, and it is not satisfied
for the following two familiar sources in the two stage case.
\begin{itemize}
\item The Gaussian source when the two side informations are not
statistically identical. This example is treated in more detail in the
next section.
\item Doubly-symmetric binary source (DSBS) with Hamming distortion
measure, when the first stage does not have side information. An
explicit calculation is given in Section \ref{sec:DSBS}.
\end{itemize}

A natural question arises as whether the aforementioned second
condition can be satisfied separately, and for this purpose the notion
of generalized successively refinable with side information is
defined. This notion can be used to delineate these two conditions
which result in the failure of a source being successively refinable.

\begin{definition}
A source $X$ is said to be {\em $N$-step generalized successively
refinable} with degraded side informations, i.e., $X \leftrightarrow
Y_N \leftrightarrow Y_{N-1} \leftrightarrow... \leftrightarrow Y_1$,
along the distortion vector $\vec{D}=(D_1,D_2,...,D_N)$, if
\begin{eqnarray}
(R_{HB}(D_1),R_{HB}(D_1,D_2)-R_{HB}(D_1),...,R_{HB}(D_1,D_2,...,D_N)-R_{HB}(D_1,D_2,...,D_{N-1}))\nonumber\\
\quad\in\mathcal{R}(\vec{D}).\nonumber
\end{eqnarray}
\end{definition}

The definition is limited to the degraded side information case,
because $R_{HB}(D_1,D_2,...,D_N)$ is known under this condition. The
notion of generalized successive refinability only considers whether
in order to achieve distortion $(D_1,D_2,...,D_N)$ with side
informations $(Y_1,Y_2,...,Y_N)$, a progressive encoder is as good as
an arbitrary encoder, but ignores whether
$R^*_{X|Y_m}(D_m)=R_{HB}(D_1,D_2,...,D_m)$ is true.

The following theorem makes explicit the connection between strictly successive refinability and the generalized version.

\begin{theorem}
\label{theorem:tworefine}
A source $X$ is $N$-step strictly successively refinable with degraded
side information along the distortion vector
$\vec{D}=(D_1,D_2,...,D_N)$, if and only if it is $N$-step generalized
successively refinable, and
$R_{HB}(D_1,D_2,...,D_m)=R^*_{X|Y_m}(D_m)$ for each $1\leq m\leq N$.
\end{theorem}

\noindent{\em Proof of Theorem \ref{theorem:tworefine}}

The sufficiency is trivial, and we only prove the necessity.  By
definition, we have
\begin{eqnarray}
\vec{r}^*=(R^*_{X|Y_1}(D_1),R^*_{X|Y_2}(D_2)-R^*_{X|Y_1}(D_1),...,R^*_{X|Y_N}(D_N)-R^*_{X|Y_{N-1}}(D_{N-1}))\in
\mathcal{R}(\vec{D}).
\end{eqnarray}
Since $\vec{r}^*$ is achievable, it must satisfy the following lower bound:
\begin{eqnarray}
\sum_{i=1}^m r_i^* \geq R_{HB}(D_1,D_2,...,D_m), \quad 1\leq m\leq N.
\end{eqnarray}
Define the rate vector
\begin{align} \vec{r}=
(R_{HB}(D_1),R_{HB}(D_1,D_2)-R_{HB}(D_1),...,R_{HB}(D_1,D_2,...,D_N)-R_{HB}(D_1,D_2,...,D_{N-1}))
\end{align}
then it follows
\begin{align}
\sum_{i=1}^m r_i=R_{HB}(D_1,D_2,...,D_m)\geq
R^*_{X|Y_m}(D_m)=\sum_{i=1}^m r_i^*\geq R_{HB}(D_1,D_2,...,D_m), \quad
1\leq m\leq N.
\end{align}
Thus the inequalities must be equality which gives
$R_{HB}(D_1,D_2,...,D_m)=R^*_{X|Y_m}(D_m)$ for $1\leq m\leq N$.  The
sum-incremental property of the rate-distortion region
$\mathcal{R}(\vec{D})$ further implies that
$\vec{r}\in\mathcal{R}(\vec{D})$, which completes the proof.
\hfill$\square$

The next theorem is also straightforward as a consequence of Theorem 1
and the definition of generalized successive refinability, thus the
proof is omitted.
\begin{theorem}
\label{theorem:generalized}
A discrete memoryless stochastically degraded source $X
\leftrightarrow Y_N \leftrightarrow Y_{N-1}
\leftrightarrow... \leftrightarrow Y_1$ is $N$-step generalized
successively refinable if and only if there exist random variables
$(W_1,W_2,...,W_N)$ satisfying the conditions given for
$\mathcal{R}^*(D_1,D_2,...,D_N)$ with
\begin{eqnarray}
R_{HB}(D_1,D_2,...,D_m)=\sum_{i=1}^m I(X;W_i|W_1,W_2,...,W_{i-1},Y_i), \quad 1\leq m \leq N.
\end{eqnarray}
\end{theorem}

Different from strictly successive refinability with degraded side
information in \cite{Steinbergmerhav:04} or the conventional
successive refinability without side information
\cite{EquitzCover:91}, there is no Markov condition involved. Though
somewhat surprising at the first sight, it is actually
straightforward, because for degraded side informations, the optimal
coding scheme naturally employs a progressive order. However, an
arbitrary source is not necessarily generalized successively
refinable along a distortion vector (pair), because a random variable $W_1^*$ optimal for the first
stage, is not necessarily optimal together with any $W_2$ for the
first two stages. An example is that any source that is not
successively refinable without side information, is not generalized
successively refinable if we take both the side information $Y_1$ and
$Y_2$ as constant.

With the definitions above, we will show in the next section that
though Gaussian source with different but degraded side informations
is not strictly successively refinable, it is indeed generalized
successively refinable. The reason for it to be not strictly
successively refinable is thus only due to the fact
$R_{HB}(D_1,D_2,...,D_j)>R^*_{X|Y_j}$ in these cases. Furthermore, we
will show that the same is true for the DSBS source. Unlike the conventional successive refinability without side information, when
side information is involved, many familiar sources are very likely to
be not strictly successively refinable unless the side information is
identical at all the stages; however, they are quite likely to be
generalized successively refinable.

\section{Gaussian Source with Different Side Informations}
\label{sec:Gauss}

We explore the Gaussian source with mean squared error distortion
measure in this section. The calculation will be focused on the
two-stage system, which is sufficient for the purpose of illustrating
the two kinds of successive refinability; however, it can be
generalized to any finite stages. We emphasize that this derivation is
{\em not} a trivial extension of the one in \cite{Heegardberger:85}
when $Y_1$ is a constant, and thus more details are included in
Appendix \ref{append:Gaussian}. Though all the discussions in the
previous sections are for discrete sources, the result can be
generalized to the Gaussian source using the techniques in
\cite{Gallagerbook}\cite{Wyner:78}.

We first recall the result in \cite{Heegardberger:85} for the two stage case,
\begin{eqnarray}
R_{HB}(D_1,D_2)=\min_{p(D_1,D_2)}[I(X;W_1|Y_1)+I(X;W_2|W_1,Y_2)],
\end{eqnarray}
where $p(D_1,D_2)$ is the set of all random variable
$(W_1,W_2)\in \mathcal{W}_1\times\mathcal{W}_2$ jointly distributed
with the generic random variables $(X,Y_1,Y_2)$, such that the
following conditions are satisfied: 
(1) $(W_1,W_2)\leftrightarrow X \leftrightarrow Y_2\leftrightarrow
Y_1$ is a Markov string; (2) there exist deterministic functions $f_1$ and $f_2$ such that
\begin{eqnarray*} 
\Expt\mathit{d}(X,f(W_1,Y_1))\leq D_1,\quad
\Expt\mathit{d}(X,f(W_1,W_2,Y_2))\leq D_2.
\end{eqnarray*}

\begin{figure}[tb]
  \centering 
\includegraphics[scale=0.35]{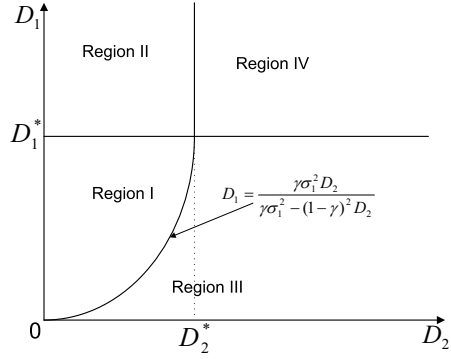}
\caption{Partition of distortion region for the quadratic Gaussian source. \label{fig:Gaussian} }
\end{figure}

The source in question is $X\sim \mathcal{N}(0,\sigma_x^2)$, i.e., a
zero mean normal random variable with variance $\sigma_x^2$. Let
$Y_1=X+N_1+N_2$ and $Y_2=X+N_2$, where $N_1\sim
\mathcal{N}(0,\sigma_1^2)$, $N_2\sim \mathcal{N}(0,\sigma_2^2)$, and
$X$, $N_1$ and $N_2$ are mutually independent and Gaussian;
further assume that $\sigma_1^2, \sigma_2^2>0$.  To facilitate the
discussions, we partition the distortion regions into the following
subregions\footnote{To make the definition of the regions to be
consistent with those in \cite{Kerpez:87}, we label the horizontal
axis as $D_2$. This convention is also used in the next section.}, as
illustrated in Fig. \ref{fig:Gaussian}, where $D_1^*$, $D_2^*$ and
$\gamma$ are defined as
\begin{eqnarray*}
D_1^*\stackrel{\Delta}{=}\frac{\sigma_x^2(\sigma_1^2+\sigma_2^2)}{\sigma_x^2+\sigma_1^2+\sigma_2^2}, \quad 
D_2^*\stackrel{\Delta}{=}\frac{\sigma_x^2\sigma_2^2}{\sigma_x^2+\sigma_2^2},\quad
\gamma\stackrel{\Delta}{=} \frac{\sigma_2^2}{\sigma_1^2+\sigma_2^2},
\end{eqnarray*}
where it is clear that $D_1^*$ and $D_2^*$ are the variance of the best MMSE linear estimator
of $X$ given $Y_1$ and $Y_2$, respectively.

The regions can be understood as follows
\begin{itemize}
\item Region I: $0<D_1\leq D_1^*$, $0<D_2\leq D_2^*$ and $D_1\geq
\frac{\gamma\sigma_1^2D_2}{\gamma\sigma_1^2-(1-\gamma)^2D_2}$. In this
region both constraints are effective.
\item Region II: $D_1> D_1^*$, $0<D_2< D_2^*$. In this region, the
first stage does not have to encode, and the problem degenerates to
Wyner-Ziv coding only for the second stage, i.e., $R_1\geq0$ and
$R_1+R_2\geq R^*_{X|Y_2}(D_2)$.
\item Region III: $D_1\leq D_1^*$ and $0<D_1<
\frac{\gamma\sigma_1^2D_2}{\gamma\sigma_1^2-(1-\gamma)^2D_2}$. In this
region, the second stage does not have to encode, and the problem
degenerates to Wyner-Ziv coding only for the first stage, i.e.,
$R_1\geq R^*_{X|Y_1}(D_1)$ and $R_2\geq 0$.
\item Region IV: $D_1> D_1^*$ and $D_2> D_2^*$. This can be achieved with zero rate,
since the side-informations are enough to satisfy the distortion constraints.
\end{itemize}

Region I is the only non-degenerate case among the four. In fact, for
any distortion pairs $(D_1,D_2)$ in Region II, III or IV, there is a
distortion pair $(D_1',D_2')$ on the boundary of Region I that
strictly improves over $(D_1,D_2)$, and is achievable using the same
rates; i.e., $R(D_1,D_2)=R(D_1',D_2')$, and $D_1\geq D_1'$, $D_2\geq
D_2'$, where at least one of inequalities holds strictly.  Since
Region I is the only non-degenerate case, it will be our focus.  For
the first stage, an obvious lower bound is the Wyner-Ziv rate
distortion function, which gives
\begin{eqnarray}
R_1\geq \frac{1}{2}\log\frac{\sigma_x^2(\sigma_1^2+\sigma_2^2)}{D_1(\sigma_x^2+\sigma_1^2+\sigma_2^2)}.
\end{eqnarray}
Using $R_{HB}(D_1,D_2)$ as the lower bound on the sum rate, we have
\begin{eqnarray}
R_1+R_2\geq R_{HB}(D_1,D_2)=\frac{1}{2}\log
\frac{\sigma_x^2\sigma_1^2\sigma_2^2}{D_2(\sigma_x^2+\sigma_1^2+\sigma_2^2)((1-\gamma)^2D_1+\gamma
\sigma_1^2)}
\label{eqn:sumrateGaussian}
\end{eqnarray}
for which the rate distortion function $R_{HB}(D_1,D_2)$ is proved in Appendix \ref{append:Gaussian}.

Not surprisingly, the following pair of random variables actually
achieve the lower bounds on $R_1$ and $R_1+R_2$ simultaneously in
Region I:
\begin{eqnarray*}
W_1=X+Z_1+Z_2, \quad
W_2=X+Z_2
\end{eqnarray*}
where $Z_1$, $Z_2$ are mutually independent zero-mean Gaussian random
variable, and independent of $(X,N_1,N_2)$, with proper choice of
variances determined by
$D_1,D_2,\sigma_1^2,\sigma_2^2,\sigma_x^2$. Alternatively, it is
obvious that this choice of $W_1$ and $W_2$ makes all the inequalities
in the lower bounding derivation satisfied with equality, thus
achieves the lower bound.

From the above discussion, it is clear that this choice of $W_1$ and
$W_2$ satisfies the condition of Theorem \ref{theorem:generalized},
and thus Gaussian source is indeed generalized successively
refinable. However, in the interior of Region I, $R_{HB}(D_1,D_2)$ is
strictly larger than $R^*_{X|Y_2}(D_2)$, which implies Gaussian source
is not successively refinable in the strict sense for these distortion
pairs by Theorem \ref{theorem:tworefine}. On the boundary between
Region I and II, as well in Region II,
$R_{HB}(D_1,D_2)=R^*_{X|Y_2}(D_2)$, thus it is indeed successively
refinable in the strict sense for these distortion pairs; however,
this degenerate case is less interesting.

\section{The Doubly-symmetric Binary Source}
\label{sec:DSBS}
In this section we consider the following special case: $X$ is a DMS with alphabet  
in $\{0,1\}$, and $P(X=0)=P(X=1)=0.5$. Side information $Y_2=Y=X\oplus
N$, where $N$ is a Bernoulli random variable independent of everything
else with $P(N=1)=p<0.5$ and $\oplus$ stands for modulo 2 addition;
alternatively, $Y$ can be taken as the output of a binary symmetric
channel with input $X$, and crossover probability $p$. $Y_1$ is a
constant, i.e., there is no side information at the first stage. The
distortion measure is the Hamming distortion $d(x,\hat{x})=x\oplus
\hat{x}$, where $\oplus$ is modulo 2 summation.

As in the Gaussian case, the function $R_{HB}(D_1,D_2)$ plays a
significant role for this source. We digress here to give a brief
review of this particular problem. The DSBS source, which is probably
the simplest discrete source in the side information scenario,
provided considerable insight into the Wyner-Ziv problem
\cite{Wynerziv:76}. Somewhat surprisingly, an explicit calculation of
$R_{HB}(D_1,D_2)$ was not found for this source. Heegard and Berger
postulated a forward test channel in \cite{Heegardberger:85}, which
was later shown to be not optimal by Kerpez \cite{Kerpez:87}. Kerpez
provided upper and lower bounds, neither of which are tight. Fleming
and Effros \cite{FlemingEffros:03} also contributed to this problem by
considering it as a rate distortion problem with mixed types of side
information. An algorithm to compute the rate-distortion function
numerically was further devised in \cite{Flemingthesis}. However an
explicit expression of the rate distortion function for this source,
and more importantly the corresponding optimal forward test channel structure have not been
given in the literature. In the process of considering our problem for
the DSBS case, we give an explicit solution to the Heegard-Berger problem as
well.

In this section we first explicitly calculate $R_{HB}(D_1,D_2)$, and
then apply the result to the successive refinement coding case, where
it will be shown that the DSBS is indeed generalized successively
refinable.

\subsection{$R_{HB}(D_1,D_2)$ for the DSBS source}

\begin{figure}[tb]
  \centering
    \includegraphics[width=6.5cm]{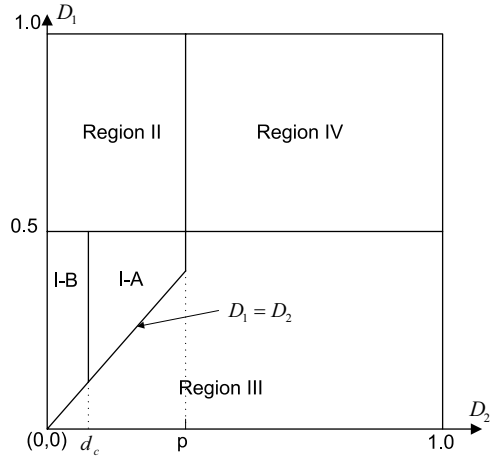}
\caption{The four parts of the rate-distortion regions. $d_c$ is the
critical distortion defined in \cite{Wynerziv:76}\label{fig:regions}}
\end{figure}

As in the Gaussian case considered in Section \ref{sec:Gauss}, it was shown in
\cite{Kerpez:87}\footnote{Note that the constraints $D_1$ and $D_2$,
which are the first and second stage distortions here, correspond to
$D_2$ and $D_1$ defined in \cite{Kerpez:87} respectively. } that the
rate distortion region can be partitioned into four subregions, three
of which are degenerate (see Fig. \ref{fig:regions}).
\begin{itemize}
\item Region I: $0\leq D_1< 0.5$ and $0\leq D_2< \min(D_1,p)$. In this
region $R(D_1,D_2)$ is a function of both $D_1$ and $D_2$, and it is
the only non-degenerate case;
\item Region II: $D_1\geq 0.5$ and $0\leq D_2\leq p$. Here the first stage does not
have to encode and therefore the problem degenerates to Wyner-Ziv encoding for
the second stage.
\item Region III:  $0\leq D_1\leq 0.5$ and $D_2\geq \min(D_1,p)$. Here the second stage
does not have to encode and hence the problem degenerates to the rate-distortion encoding 
for the first stage.
\item Region IV: $D_1> 0.5$ and $D_2> p$. Clearly the rate is zero since the distortion
constraints are trivially met.
\end{itemize}

We will need the following function from \cite{Wynerziv:76}, defined on the domain $0\leq u\leq 1$,
\begin{eqnarray*}
G(u)=h(p*u)-h(u),
\end{eqnarray*}
where $h(u)$ is the binary entropy function $h(u)=-u \log u-(1-u)\log
(1-u)$ and $u*v$ is the binary convolution for $0\leq u,v\leq 1$ and
$u*v=u(1-v)+v(1-u)$.  We will be interested only in the case $0\leq p<
0.5$. It was shown in \cite{Wynerziv:76} that $G(u)$ is (strictly)
convex; furthermore, it is easy to show that $G(u)$ is symmetric about
0.5, and is monotonically decreasing for $0\leq u\leq 0.5$; the
minimum of $G(u)$ is zero when $u=0.5$. It was also shown\footnote{In
\cite{Wynerziv:76}, the minimization was given instead as an infimum
with the feasible range of $0\leq\beta'<p$, but it can be shown that
for $D_2<p$, these two forms are equivalent.} in \cite{Wynerziv:76}
that for $0\leq D<p$
\begin{eqnarray}
\label{eqn:wzBin}
R^*_{X|Y}(D)=\min_{(\beta,\theta):0\leq\theta\leq 1, 0\leq \beta\leq p, D=\theta\beta+(1-\theta)p} [\theta G(\beta)].
\end{eqnarray}

We next define the following function 
\begin{eqnarray}
S_{D_1}(\alpha,\beta,\theta,\theta_1)=
              1-h(D_1*p)+(\theta-\theta_1)G(\alpha)+\theta_1G(\beta)+(1-\theta)G(\gamma)\nonumber  
\end{eqnarray}
where 
\begin{eqnarray*}
\gamma=\left\{\begin{array}{ll} 
\frac{D_1-(\theta-\theta_1)(1-\alpha)-\theta_1\beta}{1-\theta} & \theta\neq 1\\
0.5 & \theta=1\\
              \end{array}\right.
\end{eqnarray*}
on the domain
\begin{eqnarray*}
0\leq \theta_1\leq \theta\leq 1, \quad 0\leq \alpha,\beta\leq p, \quad p\leq \gamma\leq 1-p.
\end{eqnarray*}
Notice that $S_{D_1}(\cdot)$ is continuous at $\theta=1$.

The following theorem characterizes the rate distortion function $R_{HB}(D_1,D_2)$ in Region I. 
\begin{theorem}
\label{theorem:binHB}
For distortion pairs $(D_1,D_2)$ in Region I:
\begin{eqnarray*}
R_{HB}(D_1,D_2)=\min S_{D_1}(\alpha,\beta,\theta,\theta_1)\stackrel{\Delta}{=}S^*(D_1,D_2),
\end{eqnarray*}
where the minimization is over the domain of $S_{D_1}(\alpha,\beta,\theta,\theta_1)$, subject to the constraint
\begin{eqnarray*}
(\theta-\theta_1)\alpha+\theta_1\beta+(1-\theta)p=D_2.
\end{eqnarray*}
\end{theorem}
This theorem is proved in Appendix \ref{append:DSBS}. One notable consequence in the proof of the forward part of this theorem, is that $W_1$ can always be taken as the output of a BSC with crossover probability $D_1$ and input X. This observation is important to determine whether this source is generalized successively refinable.

The following two corollaries are useful, and are straightforward given Theorem \ref{theorem:binHB}, which are also proved in Appendix \ref{append:DSBS}. The first corollary provides a lower 
bound for $R_{HB}(D_1,D_2)$, which is easy to compute and
usually tighter than the one given in \cite{Kerpez:87}.
\begin{corollary}
\label{corollary:bin}
For distortion pairs $(D_1,D_2)$ in Region I:
\begin{eqnarray*}
R_{HB}(D_1,D_2)\geq 1-h(D_1*p)+R^*_{X|Y}(D_2).
\end{eqnarray*}
\end{corollary}

\begin{figure}[tb]
  \centering
    \includegraphics[width=10cm]{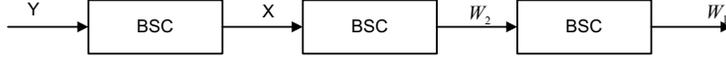}
\caption{The optimal forward test channel in Region
I-B\label{fig:special}. The crossover probability for the BSC between
$X$ and $W_2$ is $D_2$, while the crossover probability $\eta$ for the
BSC between $W_2$ and $W_1$ is such that $D_2*\eta=D_1$.}
\end{figure}

Next recall the definition of the critical distortion $d_c$ in the Wyner-Ziv problem for the DSBS source, where
\begin{eqnarray*}
\frac{G(d_c)}{d_c-p}=G'(d_c).
\end{eqnarray*}
We have the following corollary which specifies a simple forward test channel structure for the case $D_2\leq d_c$.
\begin{corollary}
\label{corollary:special}
For distortion pairs $(D_1,D_2)$ such that $D_1\leq0.5$ and $D_2\leq\min(d_c,D_1)$ (i.e., Region I-B), 
\begin{eqnarray*}
R_{HB}(D_1,D_2)=1-h(D_1*p)+G(D_2).
\end{eqnarray*}
\end{corollary}
From the proof of Corollary \ref{corollary:special}, it is seen that
the optimal forward test channel for this case is in fact a
cascade of two BSC channels depicted in Fig. \ref{fig:special}.

\subsection{Successive Refinability for the DSBS Source}
From Corollary \ref{corollary:bin}, it is evident that
$R_{HB}(D_1,D_2)>R^*_{X|Y}(D_2)$ unless $D_1=0.5$, which implies that
the DSBS is not strictly successively refinable; however, it is
generalized successively refinable. This is true because Theorem
\ref{theorem:binHB} and its proof imply that $W_1$ can always be taken
as the output of a BSC with crossover probability of $D_2$ and input
$X$. This $W_1$ and the optimal $W_2$ clearly satisfy the condition in
Theorem \ref{theorem:generalized}, thus the DSBS is indeed generalized
successively refinable.

\section{Conclusion}
\label{sec:disc}
We provided a characterization of the rate-distortion region for the
multistage successive refinement of Wyner-Ziv problem with degraded
side information, which was left open in \cite{Steinbergmerhav:04}. A
systematical comparison with the achievable region given in
\cite{Steinbergmerhav:04} was provided, and the equivalence is
established precisely. We also established a source-channel separation theorem when descriptions are transmitted over independent channels. Conditions for (strictly) successively
refinable are accordingly derived. The notion of generalized
successively refinable was introduced, in order to delineate the two
obvious factors which result in the failure of a source being
successively refinable. We showed that the Gaussian source with
multiple side informations, as well as the doubly symmetric binary
source when the first stage does not have side information, are in
fact generalized successively refinable, but not strictly successively
refinable. As such, their being not successively refinable is only due
to the uncertainty on which side information will occur, but not the
progressive encoding requirement.

\appendix
\section{Proof of the Converse of Theorem 1}
\label{appsec:AppI}

There are a total of $N$ rate constraint inequalities. We consider
bounding the rate sum $\sum_{i=1}^mR_i$ for a given $m$, where $1\leq m
\leq N$. Assume the existence of $(n,M_1,M_2,...,M_N,D_1,D_2,...,D_N)$
SR code, there exist encoding and decoding functions $\phi_i$ and
$\psi_i$ for $1\leq i \leq N$. Denote $\phi_i(X^n)$ as $T_i$. We will
use the notation $T_i^j$ to denote the vector $(T_i,T_{i+1},...,T_j)$
when $i\leq j$; if $i>j$, we take the convention that $T_i^j$ is the
empty set $\emptyset$. $(X_1,X_2,...,X_n)$ will be denoted as
$\vec{X}$ and $(Y_{j,1},Y_{j,2},...,Y_{j,n})$ as
$\vec{Y_j}$. $\vec{X}_k^-$ will be used to denote the vector
$(X_1,X_2,...,X_{k-1})$ and $\vec{X}_k^+$ to denote
$(X_{k+1},X_{k+2},...,X_{n})$. For a collection of side informations,
denote $((\vec{Y_i})_k^+,(\vec{Y_{i+1}})_k^+,...,(\vec{Y_j})_k^+)$ as
$(\vec{Y_i^j})_k^+$, and similarly for $(\vec{Y_i^j})_k^-$; they will
be combined when necessary and denoted as $(\vec{Y_i^j})_k^\pm$. The
subscript $k$ will be dropped when it is obvious from the
context. $(Y_i^j)_k$ is understood as the vector
$(Y_{i,k},Y_{i+1,k},...,Y_{j,k})$. We will assume $m>2$ such that the
quantities exist in the following proof, but it is straightforward to
verify for $m=1,2$, that the derivation degenerates in the correct
way.

The following chain of inequalities is straightforward
\begin{eqnarray}
n\sum_{i=1}^m R_i& \geq &H(T_1^m)\nonumber\\
&\geq&H(T_1^m|\vec{Y_1}) \stackrel{(a)}{=} H(T_1^m|\vec{Y_1}) - H(T_1^m|\vec{Y_1},\vec{X})\nonumber\\
&=&I(\vec{X};T_1^m|\vec{Y_1})\label{eqn:forSC}\\
&=&I(\vec{X};T_1^m\vec{Y_2^m}|\vec{Y_1})-\sum_{j=2}^mI(\vec{X};\vec{Y_j}|T_1^m\vec{Y_1^{j-1}})\label{eqn:a1}\\
&=&\sum_{k=1}^n
[I(X_k;T_1^m\vec{Y_2^m}|\vec{Y_1}\vec{X}_k^-)-\sum_{j=2}^m
I(\vec{X};Y_{j,k}|T_1^m\vec{Y_1^{j-1}}(\vec{Y_j})_k^-)]\label{eqn:a2}
\end{eqnarray}
where $(a)$ is because the index is a function of the source, 
and the last two equalities follow from the chain rule for mutual information. 
Define the term in the outer summation of (\ref{eqn:a2}) as $\Gamma_k$, i.e.,
\begin{eqnarray}
\Gamma_k=I(X_k;T_1^m\vec{Y_2^m}|\vec{Y_1}\vec{X}_k^-)-\sum_{j=2}^m I(\vec{X};Y_{j,k}|T_1^m
\vec{Y_1^{j-1}}(\vec{Y_j})_k^-)
\end{eqnarray}
For simplicity, from here on we will drop the subscript $k$ when we refer to the sequences, 
{\em e.g.,} we will denote $\vec{X}_k^-$ by $\vec{X}^-$ and $(\vec{Y_j})_k^-$ by $\vec{Y_j}^-$. 
We will work primarily with
$\Gamma_k$ until the very end of the proof. For the first term in
$\Gamma_k$
\begin{eqnarray}
I(X_k;T_1^m\vec{Y_2^m}|\vec{Y_1}\vec{X}^-)
\stackrel{(a)}{=}I(X_k;T_1^m\vec{Y_2^m}{\vec{Y^{\pm}_1}}\vec{X}^-|Y_{1,k})\geq I(X_k;T_1^m\vec{Y_2^m}{\vec{Y^{\pm}_1}}|Y_{1,k})\label{eqn:b2}
\end{eqnarray}
where (a) follows from the fact that $(X_k,Y_{1,k})$
is independent of $(\vec{X}^-,{\vec{Y^{\pm}_1}})$. Because of the
Markov string $Y_{j,k}\leftrightarrow (X_k,(Y_1^{j-1})_k)
\leftrightarrow (T_1^m\vec{X}^\pm(\vec{Y_1^{j-1}})^\pm\vec{Y_j}^-)$,
for each term in the negative summation in $\Gamma_k$, we have
\begin{eqnarray}
I(\vec{X};Y_{j,k}|T_1^m\vec{Y_1^{j-1}}{\vec{Y^-_j}})=I(X_k;Y_{j,k}|T_1^m\vec{Y_1^{j-1}}{\vec{Y^-_j}}) \label{eqn:c1}
\end{eqnarray}

Combining (\ref{eqn:b2}) and (\ref{eqn:c1}), it follows
\begin{eqnarray}
\Gamma_k&\geq&I(X_k;T_1^m\vec{Y_2^m}{\vec{Y^{\pm}_1}}|Y_{1,k})-
\sum_{j=2}^mI(X_k;Y_{j,k}|T_1^m\vec{Y_1^{j-1}}{\vec{Y^-_j}})\label{eqn:e1}
\end{eqnarray}
Applying the chain rule for the positive term in  the right hand side of (\ref{eqn:e1}), we have
\begin{eqnarray}
I(X_k;T_1^m\vec{Y_2^m}{\vec{Y^{\pm}_1}}|Y_{1,k})=
I(X_k;T_1^m\vec{Y^{\pm}_1}\vec{Y^-_2}|Y_{1,k})+
I(X_k;Y_{2,k}\vec{Y^+_2}\vec{Y_3^m}|T_1^m\vec{Y_1}\vec{Y^-_2})\label{eqn:d1}
\end{eqnarray}
For the second term in Eqn. (\ref{eqn:d1}), we have 
\begin{align}
\label{eq:SecTerm}
&I(X_k;Y_{2,k}\vec{Y^+_2}\vec{Y_3^m}|T_1^m\vec{Y_1}\vec{Y^-_2})=I(X_k;Y_{2,k}|T_1^m\vec{Y_1}\vec{Y^-_2})+I(X_k;\vec{Y_2^+}\vec{Y_3^m}|T_1^m\vec{Y_1}\vec{Y^-_2}Y_{2,k})\nonumber\\
&=I(X_k;Y_{2,k}|T_1^m\vec{Y_1}\vec{Y^-_2})+
I(X_k;\vec{Y^+_2}\vec{Y^-_3}|T_1^m\vec{Y_1}\vec{Y^-_2}Y_{2,k})
+I(X_k;Y_{3,k}\vec{Y^+_3}\vec{Y_4^m}|T_1^m\vec{Y_1^2}\vec{Y^-_3})\nonumber \\
&=I(X_k;Y_{2,k}|T_1^m\vec{Y_1}\vec{Y^-_2})+I(X_k;\vec{Y^+_2}\vec{Y^-_3}|T_1^m\vec{Y_1}\vec{Y^-_2}Y_{2,k})\nonumber\\
&\qquad\qquad\qquad+I(X_k;Y_{3,k}|T_1^m\vec{Y_1^2}\vec{Y^-_3})+I(X_k;\vec{Y^+_3}\vec{Y_4^m}|T_1^m\vec{Y_1^2}\vec{Y^-_3}Y_{3,k}).
\end{align}
Continuing this decomposition, it finally gives
\begin{align}
&I(X_k;Y_{2,k}\vec{Y^+_2}\vec{Y_3^m}|T_1^m\vec{Y_1}\vec{Y^-_2})
=\sum_{j=2}^mI(X_k;Y_{j,k}|T_1^m\vec{Y_1^{j-1}}{\vec{Y^-_j}})\nonumber\\
&\qquad\qquad+\sum_{j=2}^{m-1}I(X_k;\vec{Y_j}^+
\vec{Y^-_{j+1}}|T_1^m\vec{Y_1^{j-1}}\vec{Y^-_j}Y_{j,k})+I(X_k;\vec{Y^+_m}|T_1^m\vec{Y_1^{m-1}}\vec{Y^-_m}Y_{m,k}).
\label{eqn:e2}
\end{align}
Substituting this in (\ref{eqn:d1}), we get
\begin{align}
&&I(X_k;T_1^m\vec{Y_2^m}{\vec{Y^{\pm}_1}}|Y_{1,k})= I(X_k;T_1^m\vec{Y^{\pm}_1}\vec{Y^-_2}|Y_{1,k})
+\sum_{j=2}^mI(X_k;Y_{j,k}|T_1^m\vec{Y_1^{j-1}}{\vec{Y^-_j}})\nonumber\\
&&\quad\qquad\qquad+\sum_{j=2}^{m-1}I(X_k;\vec{Y_j}^+
\vec{Y^-_{j+1}}|T_1^m\vec{Y_1^{j-1}}\vec{Y^-_j}Y_{j,k})+I(X_k;\vec{Y^+_m}|T_1^m\vec{Y_1^{m-1}}\vec{Y^-_m}Y_{m,k}).
\label{eq:PosTerm}
\end{align}
Therefore, substituting (\ref{eq:PosTerm}) into (\ref{eqn:e1}) we see that
the negative term in (\ref{eqn:e1}) cancels out the second term on the RHS of (\ref{eq:PosTerm}), which
gives
\begin{eqnarray}
&&\Gamma_k\geq I(X_k;T_1^m\vec{Y_1}^\pm\vec{Y^-_2}|Y_{1,k})\nonumber\\
&&\quad\quad\quad+\sum_{j=2}^{m-1}I(X_k;\vec{Y^+_j}\vec{Y^-_{j+1}}|T_1^m\vec{Y_1^{j-1}}\vec{Y^-_j}Y_{j,k})+
I(X_k;\vec{Y^+_m}|T_1^m\vec{Y_1^{m-1}}\vec{Y^-_m}Y_{m,k})\label{eqn:f1}.
\end{eqnarray}

For the first term in (\ref{eqn:f1}), we have
\begin{eqnarray}
I(X_k;T_1^m\vec{Y^{\pm}_1}\vec{Y^-_2}|Y_{1,k})
=I(X_k;T_1\vec{Y^{\pm}_1}|Y_{1,k})+I(X_k;T_2^m\vec{Y^-_2}|T_1\vec{Y_1}).\label{eqn:g1}
\end{eqnarray}
We claim that 
\begin{eqnarray}
I(X_k;T_2^m\vec{Y^-_2}|T_1\vec{Y_1})\geq I(X_k;T_2^m\vec{Y^-_2}|T_1\vec{Y_1}Y_{2,k})
\label{eqn:h3}
\end{eqnarray}
and more generally for $2\leq j \leq m$
\begin{eqnarray}
I(X_k;T_j^m\vec{Y^-_j}|T_1^{j-1}\vec{Y_1^{j-1}})\geq
I(X_k;T_j^m\vec{Y^-_j}|T_1^{j-1}\vec{Y_1^{j-1}}Y_{j,k})\label{eqn:inequality}
\end{eqnarray}
which can be justified as follows
\begin{eqnarray}
&&I(X_k;T_j^m\vec{Y^-_j}|T_1^{j-1}\vec{Y_1^{j-1}})-I(X_k;T_j^m\vec{Y^-_j}|T_1^{j-1}\vec{Y_1^{j-1}}Y_{j,k})\nonumber\\
&=&H(X_k|T_1^{j-1}\vec{Y_1^{j-1}})-H(X_k|T_1^m\vec{Y_1^{j-1}}\vec{Y^-_j})\nonumber\\
&&\qquad\qquad\qquad-
H(X_k|T_1^{j-1}\vec{Y_1^{j-1}}Y_{j,k})+H(X_k|T_1^{m}\vec{Y_1^{j-1}}\vec{Y^-_j}Y_{j,k})\nonumber\\
&=&I(X_k;Y_{j,k}|T_1^{j-1}\vec{Y_1^{j-1}})-I(X_k;Y_{j,k}|T_1^m\vec{Y^-_j}\vec{Y_1^{j-1}})\nonumber\\
&=&H(Y_{j,k}|T_1^{j-1}\vec{Y_1^{j-1}})-H(Y_{j,k}|X_kT_1^{j-1}\vec{Y_1^{j-1}})\nonumber\\&&
\qquad\qquad\qquad-H(Y_{j,k}|T_1^m\vec{Y^-_j}\vec{Y_1^{j-1}})+H(Y_{j,k}|X_kT_1^m\vec{Y^-_j}\vec{Y_1^{j-1}})\nonumber\\
&\stackrel{(a)}{=}&I(Y_{j,k};T_j^m\vec{Y^-_j}|T_1^{j-1}\vec{Y_1^{j-1}})\label{eqn:h1}\geq 0
\end{eqnarray}
where (a) is due to the Markov condition
$Y_{j,k}\leftrightarrow (X_k,(Y_1^{j-1})_k) \leftrightarrow
(T_1^m\vec{Y^-_j}(\vec{Y_1^{j-1}})^\pm\vec{X}^\pm)$ implies the
reduced Markov condition $Y_{j,k}\leftrightarrow (X_k,(Y_1^{j-1})_k)
\leftrightarrow (T_1^m\vec{Y^-_j}(\vec{Y_1^{j-1}})^\pm)$. Assume for
now $m>2$, and consider the following summation of the second term in
(\ref{eqn:g1}) and the second term in (\ref{eqn:f1})
\begin{eqnarray}
&&I(X_k;T_2^m\vec{Y^-_2}|T_1\vec{Y_1})+
\sum_{j=2}^{m-1}I(X_k;\vec{Y^+_j}\vec{Y^-_{j+1}}|T_1^m\vec{Y_1^{j-1}}\vec{Y^-_j}Y_{j,k})\nonumber\\
&\stackrel{(a)}{\geq}&I(X_k;T_2^m\vec{Y^-_2}|T_1\vec{Y_1}Y_{2,k})+
\sum_{j=2}^{m-1}I(X_k;\vec{Y^+_j}\vec{Y^-_{j+1}}|T_1^m\vec{Y_1^{j-1}}\vec{Y^-_j}Y_{j,k})\nonumber\\
&=&I(X_k;T_2^m\vec{Y^-_2}|T_1\vec{Y_1}Y_{2,k})+
I(X_k;\vec{Y_2}^+\vec{Y^-_{3}}|T_1^m\vec{Y_1}\vec{Y^-_2}Y_{2,k})\nonumber\\&&\qquad\qquad\qquad\qquad\qquad\qquad\qquad\qquad\qquad+
\sum_{j=3}^{m-1}I(X_k;\vec{Y^+_j}\vec{Y^-_{j+1}}|T_1^m\vec{Y_1^{j-1}}\vec{Y^-_j}Y_{j,k})\nonumber\\
&\stackrel{(b)}{=}&I(X_k;T_2^m\vec{Y_2}^\pm\vec{Y^-_3}|T_1\vec{Y_1}Y_{2,k})+
\sum_{j=3}^{m-1}I(X_k;\vec{Y_j}^+\vec{Y^-_{j+1}}|T_1^m\vec{Y_1^{j-1}}\vec{Y^-_j}Y_{j,k})\nonumber\\
&=&I(X_k;T_2\vec{Y_2}^\pm|T_1\vec{Y_1}Y_{2,k})+I(X_k;T_3^m\vec{Y^-_3}|T_1^2\vec{Y_1^2})+
\sum_{j=3}^{m-1}I(X_k;\vec{Y^+_j}\vec{Y^-_{j+1}}|T_1^m\vec{Y_1^{j-1}}\vec{Y^-_j}Y_{j,k}),\nonumber\\\label{eqn:h5}
\end{eqnarray}
where $(a)$ follows because of (\ref{eqn:inequality}) and $(b)$ follows due to chain rule.
Notice for the second term in (\ref{eqn:h5}), we can again apply
inequality (\ref{eqn:inequality}), and continue sequentially along
this way, which finally gives
\begin{eqnarray}
&&I(X_k;T_2^m\vec{Y^-_2}|T_1\vec{Y_1})+
\sum_{j=2}^{m-1}I(X_k;\vec{Y^+_j}\vec{Y^-_{j+1}}|T_1^m\vec{Y_1^{j-1}}\vec{Y^-_j}Y_{j,k})\nonumber\\
&&\qquad\geq\sum_{j=2}^{m-1}I(X_k;T_j\vec{Y^{\pm}_j}|T_1^{j-1}\vec{Y_1^{j-1}}Y_{j,k})+I(X_k;T_m\vec{Y^-_m}|T_1^{m-1}\vec{Y_1^{m-1}})\label{eqn:h7}
\end{eqnarray}
Combining (\ref{eqn:f1}), (\ref{eqn:g1}) and (\ref{eqn:h7}) gives
\begin{eqnarray}
\Gamma_k&\geq&I(X_k;T_1\vec{Y^{\pm}_1}|Y_{1,k})+
\sum_{j=2}^{m-1}I(X_k;T_j\vec{Y^{\pm}_j}|T_1^{j-1}\vec{Y_1^{j-1}}Y_{j,k})\nonumber\\
&&+I(X_k;T_m\vec{Y^-_m}|T_1^{m-1}\vec{Y_1^{m-1}})+
I(X_k;\vec{Y^+_m}|T_1^m\vec{Y_1^{m-1}}\vec{Y^-_m}Y_{m,k})\label{eqn:h9}\\
&\geq&\sum_{j=1}^{m}I(X_k;T_j\vec{Y^{\pm}_j}|T_1^{j-1}\vec{Y_1^{j-1}}Y_{j,k})\label{eqn:h8}.
\end{eqnarray}
where inequality (\ref{eqn:inequality}) is applied on the third term
in (\ref{eqn:h9}). It is straightforward to verify that inequality
(\ref{eqn:h8}) is still valid if $m=1$ or $m=2$, when the proper
convention of empty set is taken.

In (\ref{eqn:h8}), the conditioning on $(Y_1^{j-1})_k$ has to be
removed to reach the desired form, which can indeed be done due to the
degradedness of the side informations. More precisely, for $2\leq j
\leq m$
\begin{eqnarray}
&&I(X_k;T_j\vec{Y^{\pm}_j}|T_1^{j-1}\vec{Y_1^{j-1}}Y_{j,k})-I(X_k;T_j\vec{Y^{\pm}_j}|T_1^{j-1}(\vec{Y_1^{j-1}})^\pm
Y_{j,k})\nonumber\\
&=&H(X_k|T_1^{j-1}\vec{Y_1^{j-1}}Y_{j,k})-H(X_k|T_1^{j}\vec{Y_1^{j}})-H(X_k|T_1^{j-1}(\vec{Y_1^{j-1}})^\pm
Y_{j,k})+H(X_k|T_1^{j}(\vec{Y_1^{j}})^\pm Y_{j,k})\nonumber\\
&=&-I(X_k;(Y_1^{j-1})_k|T_1^{j-1}(\vec{Y_1^{j-1}})^\pm
Y_{j,k})+I(X_k;(Y_1^{j-1})_k|T_1^{j}(\vec{Y_1^{j}})^\pm
Y_{j,k})=0\label{eqn:j1}
\end{eqnarray}
where in fact both the terms in the (\ref{eqn:j1}) are zero, due to
the Markov condition $(Y_1^{j-1})_k\leftrightarrow Y_{j,k}
\leftrightarrow (\vec{X}T_1^m(\vec{Y_1^m})^\pm)$ implies the reduced
Markov condition $(Y_1^{j-1})_k\leftrightarrow Y_{j,k} \leftrightarrow
(X_kT_1^j(\vec{Y_1^{j}})^\pm)$. Thus we reach the form
\begin{eqnarray}
\Gamma_k&\geq&\sum_{j=1}^{m}I(X_k;T_j\vec{Y_j}^\pm|T_1^{j-1}(\vec{Y_1^{j-1}})^\pm
Y_{j,k})=\sum_{j=1}^{m}I(X_k;T_1^j\vec{Y_j}^\pm|T_1^{j-1}(\vec{Y_1^{j-1}})^\pm
Y_{j,k})\label{eq:GammaForm}.
\end{eqnarray}
Define $W_{j,k}=(T_1^j,(\vec{Y}_j)_k^\pm)$ and by substituting (\ref{eq:GammaForm})
into (\ref{eqn:a2}) we have for $1\leq m \leq N$,
\begin{eqnarray}
n\sum_{i=1}^m R_i\geq \sum_{k=1}^n\sum_{j=1}^mI(X_k;W_{j,k}|(W_1^{j-1})_k,Y_{j,k})\label{eqn:final}
\end{eqnarray}
Therefore the Markov condition
$(W_{1,k},W_{2,k},...,W_{N,k})\leftrightarrow X_k \leftrightarrow
Y_{N,k} \leftrightarrow Y_{N-1,k}\leftrightarrow...\leftrightarrow
Y_{1,k}$ is true. Next introduce the time sharing random variable $Q$,
which is independent of the multisource, and uniformly distributed
over $I_n$. Define $W_j=(W_{j,Q},Q)$. The existence of function $f_j$
follows by defining
\begin{eqnarray}
f_j(W_j,Y_j)=\psi_{j,Q}(\phi_1(\vec{X}),\phi_2(\vec{X}),...,\phi_j(\vec{X}),\vec{\vec{Y_j}})
\end{eqnarray}
because $W_j$ includes $T_1^j\vec{Y_j}^\pm$, which leads to the
fulfillment of the distortion constraint
\begin{eqnarray}
\Expt
\mathit{d}(X,f_j(W_j,Y_j))=\frac{1}{n}\sum_{i=1}^n
\Expt\mathit{d}(X_i,\psi_{j,i}(\phi_1(\vec{X}),\phi_2(\vec{X}),...,\phi_j(\vec{X}),\vec{Y_j}))\leq
D_j,\quad  1\leq j \leq N\nonumber\\
\end{eqnarray}
and the Markov condition $(W_1,W_2,...,W_N)\leftrightarrow X
\leftrightarrow Y_N \leftrightarrow
Y_{N-1}\leftrightarrow...\leftrightarrow Y_1$ is still true.  It only
remains to show the bound (\ref{eqn:final}) can be writen in single
letter form in $W_j$, but this is straightforward following the
approach on pg. 435 of \cite{CoverThomas} (see also
\cite{Steinbergmerhav:04}). The bounds on the alphabet size is by applying conventional argument (see \cite{CsiszarKorner}). This completes the proof. \hfill$\square$

\section{Proof of Theorem 2}
\label{append:SCcoding}

The forward part is trivially implied by Theorem 1 and the conventional channel coding theorem, and thus we only give an outline of the converse part.

By Lemma 8.9.2 in \cite{CoverThomas}, we have
\begin{eqnarray}
\label{eqn:892}
n\sum_{i=1}^m \rho_iC_i \geq \sum_{i=1}^m I(X^{n_i}_{c,i};Y^{n_i}_{c,i})
\end{eqnarray}
where $n_i=n\rho_i$, and $\rho_i$ is the number of channel use per source symbol for the $i$-th channel. Notice that 
\begin{eqnarray}
&&I(X^{n_1}_{c,1}X^{n_2}_{c,2},...,X^{n_m}_{c,m};Y^{n_1}_{c,1}Y^{n_2}_{c,2},...,Y^{n_m}_{c,m})\nonumber\\
&\stackrel{(a)}{=}&I(X^{n_1}_{c,1}X^{n_2}_{c,2},...,X^{n_m}_{c,m};Y^{n_1}_{c,1})+I(X^{n_1}_{c,1}X^{n_2}_{c,2},...,X^{n_m}_{c,m};Y^{n_2}_{c,2}Y^{n_3}_{c,3},...,Y^{n_m}_{c,m}|Y^{n_1}_{c,1})\nonumber\\
&\stackrel{(b)}{=}&I(X^{n_1}_{c,1};Y^{n_1}_{c,1})+I(X^{n_1}_{c,1}X^{n_2}_{c,2},...,X^{n_m}_{c,m};Y^{n_2}_{c,2}Y^{n_3}_{c,3},...,Y^{n_m}_{c,m}|Y^{n_1}_{c,1})\nonumber\\
&=&I(X^{n_1}_{c,1};Y^{n_1}_{c,1})+H(Y^{n_2}_{c,2}Y^{n_3}_{c,3},...,Y^{n_m}_{c,m}|Y^{n_1}_{c,1})-H(Y^{n_2}_{c,2}Y^{n_3}_{c,3},...,Y^{n_m}_{c,m}|Y^{n_1}_{c,1}X^{n_1}_{c,1}X^{n_2}_{c,2},...,X^{n_m}_{c,m})\nonumber\\
&\stackrel{(c)}{=}&I(X^{n_1}_{c,1};Y^{n_1}_{c,1})+H(Y^{n_2}_{c,2}Y^{n_3}_{c,3},...,Y^{n_m}_{c,m}|Y^{n_1}_{c,1})-H(Y^{n_2}_{c,2}Y^{n_3}_{c,3},...,Y^{n_m}_{c,m}|X^{n_2}_{c,2},...,X^{n_m}_{c,m})\nonumber\\
&\stackrel{(d)}{\leq}&I(X^{n_1}_{c,1};Y^{n_1}_{c,1})+H(Y^{n_2}_{c,2}Y^{n_3}_{c,3},...,Y^{n_m}_{c,m})-H(Y^{n_2}_{c,2}Y^{n_3}_{c,3},...,Y^{n_m}_{c,m}|X^{n_2}_{c,2},...,X^{n_m}_{c,m})\nonumber\\
&=&I(X^{n_1}_{c,1};Y^{n_1}_{c,1})+I(Y^{n_2}_{c,2}Y^{n_3}_{c,3},...,Y^{n_m}_{c,m};X^{n_2}_{c,2},...,X^{n_m}_{c,m})
\end{eqnarray}
where (a) is by chain rule, and (b) and (c) are because the channels are independent, i.e.,
\begin{eqnarray} P_{Y_{c,1}Y_{c,2},...,Y_{c,m}|X_{c,1}X_{c,2},...,X_{c,m}}=P_{Y_{c,1}|X_{c,1}}P_{Y_{c,2}|X_{c,2}}...P_{Y_{c,m}|X_{c,m}}
\end{eqnarray}
which implies the Markov conditions $\{X_{c,j}\}_{j\neq i}\leftrightarrow X_{c,i} \leftrightarrow Y_{c,i}$ and $\{Y_{c,j}\}_{j\neq i} \leftrightarrow \{X_{c,j}\}_{j\neq i} \leftrightarrow (X_{c,i},Y_{c,i})$; (d) is because conditioning reduces entropy.

Continue this decomposition and combine it with (\ref{eqn:892}), we have
\begin{eqnarray}
n\sum_{i=1}^m \rho_iC_i &\geq& \sum_{i=1}^m I(X^{n_i}_{c,i};Y^{n_i}_{c,i}) \geq I(X^{n_1}_{c,1}X^{n_2}_{c,2},...,X^{n_m}_{c,m};Y^{n_1}_{c,1}Y^{n_2}_{c,2},...,Y^{n_m}_{c,m})\nonumber\\
&\stackrel{(a)}{\geq}& I(X^n;Y^{n_1}_{c,1}Y^{n_2}_{c,2},...,Y^{n_m}_{c,m})\nonumber\\
&\stackrel{(b)}{=}&I(X^nY_1^n;Y^{n_1}_{c,1}Y^{n_2}_{c,2},...,Y^{n_m}_{c,m})\nonumber\\
&=&I(Y_1^n;Y^{n_1}_{c,1}Y^{n_2}_{c,2},...,Y^{n_m}_{c,m})+I(X^n;Y^{n_1}_{c,1}Y^{n_2}_{c,2},...,Y^{n_m}_{c,m}|Y_1^n)\nonumber\\
&\geq&I(X^n;Y^{n_1}_{c,1}Y^{n_2}_{c,2},...,Y^{n_m}_{c,m}|Y_1^n) \label{eqn:SCfinal}
\end{eqnarray}
where (a) is due to data processing inequality, and (b) because the Markov chain $Y_1 \leftrightarrow X\leftrightarrow Y_{c,i}$. At this point the similarity between (\ref{eqn:SCfinal}) and (\ref{eqn:forSC}) is quite clear. Using the same steps as in the derivation as in the proof of Theorem 1, the converse of Theorem 2 is proved. \hfill$\square$

\section{Proof of Theorem 3}
\label{appsec:AppII}

We first prove for the special case $N=2$ without invoking Theorem 1
that $\mathcal{R}^*(\vec{D})=\mathcal{R}(\vec{D})$. The proof of
Theorem 3 then follows from invoking Theorem 1 for one direction and
extending the proof of $N=2$ for the other direction.

\noindent{\em Proof for the case of $N=2$}

We first prove that $\hat{\mathcal{R}}_2^*(\vec{D}) \subseteq
\mathcal{R}_2^*(\vec{D})$, where the subscript 2 stands for $N=2$. For
an arbitrary rate pair $(r_1,r_2)\in\hat{\mathcal{R}}_2^*(D_1,D_2)$,
there exist $3$ random variables $V_{1,1},V_{1,2}$ and $V_{2,2}$, and
the corresponding functions $f_1(V_{1,1},Y_1)$ and $f_2(V_{2,2},Y_2)$,
such that
\begin{eqnarray}
r_1&\geq&I(X;V_{1,1}|Y_1)+I(X;V_{1,2}|V_{1,1},Y_2)\label{eqn:k51}\\
r_2&\geq&I(X;V_{2,2}|V_{1,1},V_{1,2},Y_2)
\label{eqn:k5}
\end{eqnarray}
and the distortion constraints are satisfied. Inequalities (\ref{eqn:k51}) and (\ref{eqn:k5}) imply that 
\begin{eqnarray*}
r_1 &\geq& I(X;V_{1,1}|Y_1)\\ 
r_1+r_2 &\geq& I(X;V_{1,1}|Y_1)+I(X;V_{1,2}|V_{1,1},Y_2)+I(X;V_{2,2}|V_{1,1},V_{1,2},Y_2)\nonumber\\
&=&I(X;V_{1,1}|Y_1)+I(X;V_{1,2},V_{2,2}|V_{1,1},Y_2)
\end{eqnarray*}
Now define $W_{1}=V_{1,1}$ and $W_2=(V_{1,1},V_{1,2})$, and it follows that
\begin{eqnarray*}
r_1 &\geq& I(X;W_1|Y_1)\\ 
r_1+r_2 &\geq& I(X;W_1|Y_1)+I(X;W_2|W_1,Y_2)
\end{eqnarray*}
and $(W_1,W_2)$ is a pair of random variables satisfying the condition
for $\mathcal{R}_2^*(D_1,D_2)$ and thus $(r_1,r_2)\in
\mathcal{R}_2^*(D_1,D_2)$, which shows that
$\hat{\mathcal{R}}_2^*(\vec{D}) \subseteq \mathcal{R}_2^*(\vec{D})$ since
trivially the distortion constraints are also met.

To prove the other direction, i.e., $\hat{\mathcal{R}}_2^*(D_1,D_2)
\supseteq \mathcal{R}_2^*(D_1,D_2)$, assume
$(r_1,r_2)\in\mathcal{R}_2^*(D_1,D_2)$. There exist random variables
$W_1$ and $W_2$, and two corresponding functions $f_1(W_1,Y_1)$ and
$f_2(W_2,Y_2)$, such that
\begin{eqnarray}
r_1&\geq& I(X;W_1|Y_1)\\
r_1+r_2&\geq& I(X;W_1|Y_1)+I(X;W_2|W_1,Y_2)
\end{eqnarray}
and the distortion constraints are met. Let $\Delta r_1=r_1-I(X;W_1|Y_1)$. 
We claim that for any $0\leq \Delta r_1\leq I(X;W_2|W_1,Y_2)$, there exists a random variable $V$, such that 
\begin{eqnarray}
&&\Delta r_1=I(X;V|W_1,Y_2)\\
&&I(X;V|W_1,Y_2)+I(X;W_2|W_1,V,Y_2)=I(X;W_2|W_1,Y_2).
\end{eqnarray}
There are many ways to construct $V$, for example we can construct
$V=(W_2(J),J)$, where $J$ is a Bernoulli random variable independent
of everything else with $p(J=1)=u$; when $J=1$, $W_2(J)=W_2$ and $W_2(J)$ is a
fixed constant otherwise; $I(X;V|W_1,Y_2)$ can be any real value in
the interval $[0,I(X;W_2|W_1,Y_2)]$ by choosing $u$ appropriately.
For a more thorough treatment on this topic in the context of rate
splitting in multiple access channel, see \cite{Rimoldiurbanke:01}. It
follows that for this case
\begin{eqnarray}
r_1&=& I(X;W_1|Y_1)+I(X;V|W_1,Y_2)\\
r_2&\geq& I(X;W_1|Y_1)+I(X;W_2|W_1,Y_2)-r_1\nonumber\\
&=&I(X;W_2|W_1,V,Y).
\end{eqnarray}
Now define $V_{1,1}=W_1$, $V_{1,2}=V$ and $V_{2,2}=W_2$. The random
variables $(V_{1,1},V_{1,2},V_{2,2})$ clearly satisfy the definition
given for $\hat{\mathcal{R}}^*(D_1,D_2)$, and thus
$(r_1,r_2)\in\hat{\mathcal{R}}^*(D_1,D_2)$ for this case.  On the
other hand, if $\Delta r_1\geq I(X;W_2|W_1,Y_2)$, then defines
$V_{1,1}=W_1$, $V_{1,2}=W_2$ and $V_{2,2}=W_2$. The non-negativity
condition $r_2\geq0$ implies $r_2\geq
I(X;V_{2,2}|V_{1,1},V_{1,2},Y_2)$.  Since the reconstruction functions
$f_1(W_1,Y_1)=f_1(V_{1,1},Y_1)$ and $f_2(W_2,Y_2)=f_2(V_{2,2},Y_2)$ 
satisfy the distortion constraints, the proof is completed. \hfill$\square$

\noindent{\em Proof of Theorem 3}

Since $\hat{\mathcal{R}}^*(\vec{D})$ is an achievable region, we have
trivially $\hat{\mathcal{R}}^*(\vec{D})\subseteq
\mathcal{R}(\vec{D})=\mathcal{R}^*(\vec{D})$ due to Theorem 1. For the
inclusion of the other direction, the proof for the case $N=2$ can
clearly be extended straightforwardly, by sequentially constructing
random variable corresponding to $\{V_{i,j}\},j\geq i$. This completes the proof for Theorem
3. \hfill$\square$

\section{Lower Bound on the Sum-rate for the Gaussian Source}
\label{append:Gaussian}

To lower bound the sum-rate to achieve $(D_1,D_2)$ with side information $(Y_1,Y_2)$, consider the following quantity,
\begin{eqnarray} 
&&I(X;W_1|Y_1)+I(X;W_2|W_1,Y_2)\nonumber\\
&=&H(X|Y_1)-H(X|W_1,Y_1)+H(X|W_1,Y_2)-H(X|W_1,W_2,Y_2)\nonumber\\
&\stackrel{(a)}{=}&H(X|Y_1)-H(X|W_1,W_2,Y_2)-I(X;Y_2|W_1,Y_1)\label{eqn:k6}\\
&\stackrel{(b)}{=}&H(X|Y_1)-H(X|W_1,W_2,Y_2)-H(Y_2|W_1,Y_1)+H(Y_2|X,Y_1)\label{eqn:k9}
\end{eqnarray}
where we can see $(a)$ follows since $I(W_1,X;Y_1|Y_2)=I(W_1;Y_1|Y_2)+I(X;Y_1|W_1,Y_2)=0$
due to the Markov condition $W_1\leftrightarrow
X\leftrightarrow Y_2\leftrightarrow Y_1$, which implies
that $I(X;Y_1|W_1,Y_2)=H(X|W_1,Y_2)-H(X|W_1,Y_2,Y_1)=0$. In an identical 
manner $(b)$ is due to, $I(W_1;Y_2|X,Y_1)=H(Y_2|X,Y_1)-H(Y_2|X,Y_1,W_1)=0$. The quantities
$H(X|Y_1)$ and $H(Y_2|X,Y_1)$ are only dependent on the
multi-source. We bound the second term in (\ref{eqn:k9}) as follows
\begin{eqnarray}
H(X|W_1,W_2,Y_2)&=& H(X-\Expt(X|W_1,W_2,Y_2)|W_1,W_2,Y_2)\nonumber\\
&\leq&H(X-\Expt(X|W_1,W_2,Y_2))\nonumber\\
&\leq&H(\mathcal{N}(0,\Expt(X-\Expt(X|W_1,W_2,Y_2))^2))\label{eqn:k7}\\
&\leq&\frac{1}{2}\log(2\pi e D_2)\label{eqn:k8}
\end{eqnarray}
where in (\ref{eqn:k7}) we use the fact that normal distribution
maximizes the entropy for a given second moment, and in (\ref{eqn:k8})
the fact that the variance of $\Expt(X-\Expt(X|W_1,W_2,Y_2))^2\leq
D_2$ because of the existence of function $f_2(W_1,W_2,Y_2)$ to
reconstruct $X$ with distortion $D_2$.

To bound the third term in (\ref{eqn:k9}), write $Y_2=X+N_2$ as follows
\begin{eqnarray*}
X+N_2&=&X+N_2+\frac{\sigma_2^2}{\sigma_1^2+\sigma_2^2}(N_1+N_2)-\frac{\sigma_2^2}{\sigma_1^2+\sigma_2^2}(N_1+N_2)\\
&=&\frac{\sigma_2^2}{\sigma_1^2+\sigma_2^2}(X+N_1+N_2)+
\frac{\sigma_1^2}{\sigma_1^2+\sigma_2^2}X+[N_2-\frac{\sigma_2^2}{\sigma_1^2+\sigma_2^2}(N_1+N_2)]\nonumber\\
&=&\gamma Y_1+(1-\gamma)X+[(1-\gamma)N_2-\gamma N_1],
\end{eqnarray*}
where $\gamma=\frac{\sigma_2^2}{\sigma^2_1+\sigma^2_2}$ as in Section
\ref{sec:Gauss}.  It can be seen that $[(1-\gamma)N_2-\gamma N_1]$ is
independent of $Y_1$, by checking the fact
$\Expt(Y_1[(1-\gamma)N_2-\gamma N_1])=0$ and recalling that they are
jointly zero-mean Gaussian. Further notice $X$ is independent of
$(N_1,N_2)$, which implies $[(1-\gamma)N_2-\gamma N_1]$ is also
independent of $W_1$.  Thus we have
\begin{eqnarray}
H(Y_2|W_1,Y_1)&=& H(\gamma Y_1+(1-\gamma)X+[(1-\gamma)N_2-\gamma
N_1]|W_1,Y_1)\nonumber\\ &=&H((1-\gamma)X+[(1-\gamma)N_2-\gamma
N_1]|W_1,Y_1)\nonumber\\
&=&H((1-\gamma)[X-\Expt(X|W_1,Y_1)]+[(1-\gamma)N_2-\gamma
N_1]|W_1,Y_1)\nonumber\\
&\leq&H((1-\gamma)[X-\Expt(X|W_1,Y_1)]+[(1-\gamma)N_2-\gamma
N_1])\nonumber\\
&\leq&H(\mathcal{N}(0,\Expt\{(1-\gamma)[X-\Expt(X|W_1,Y_1)]+[(1-\gamma)N_2-\gamma
N_1]\}^2))\\
&\leq&H(\mathcal{N}(0,(1-\gamma)^2D_1+(1-\gamma)^2\sigma_2^2+\gamma^2\sigma_1^2))\label{eqn:indp}\\
&=&\frac{1}{2}\log[2\pi
e((1-\gamma)^2D_1+(1-\gamma)^2\sigma_2^2+\gamma^2\sigma_1^2))]\nonumber\\
&=&\frac{1}{2}\log[2\pi e((1-\gamma)^2D_1+\gamma \sigma_1^2))]
\label{eq:2term}
\end{eqnarray}
where in (\ref{eqn:indp}), we used the fact that
$[X-\Expt(X|W_1,Y_1)]$ is independent of $[(1-\gamma)N_2-\gamma
N_1]$. Using (\ref{eqn:k8}) and (\ref{eq:2term}) in (\ref{eqn:k9}) gives
\begin{eqnarray}
R_1+R_2\geq \frac{1}{2}\log
\frac{\sigma_x^2\sigma_1^2\sigma_2^2}{D_2(\sigma_x^2+\sigma_1^2+\sigma_2^2)((1-\gamma)^2D_1+\gamma
\sigma_1^2)}
\end{eqnarray}
Note that this lower bound is only tight and achievable when both
$D_1$ and $D_2$ are effective, i.e., in Region I. When $D_2$ is not
effective, the bound that
\begin{eqnarray*}
R_1+R_2\geq R_1\geq\frac{1}{2}\log(\frac{\sigma_x^2(\sigma_1^2+\sigma_2^2)}{D_1(\sigma_x^2+\sigma_1^2+\sigma_2^2)})
\end{eqnarray*} 
is in fact achievable with equality. By comparing the above two
bounds, it can be seen that this corresponds to the condition $D_2\leq
\frac{\gamma D_1\sigma_1^2}{(1-\gamma)^2D_1+\gamma\sigma_1^2}$ or
equivalently $D_1\geq
\frac{\gamma\sigma_1^2D2}{\gamma\sigma_1^2-(1-\gamma)^2D_2}$ when
$D_2\leq D_2^*$. \hfill$\square$

\section{Proof of the Theorem and Corollaries for the DSBS}
\label{append:DSBS}

\subsection{Proof of Theorem \ref{theorem:binHB}}

We will need the following lemma from \cite{Kerpez:87} to simplify the calculation.
\begin{lemma}
For $(W_1,W_2)\in p(D_1,D_2)$
\begin{eqnarray}
I(X;W_1)+I(X;W_2|YW_1)=H(X)-H(Y|W_1)+H(Y|W_1W_2)-H(X|W_1W_2).
\end{eqnarray}
\end{lemma}

\noindent{\em The lower bound}

Let $(W_1,W_2)\in P(D_1,D_2)$ define a joint distribution with
$(X,Y)$.  Furthermore, assume the functions $f_1$ and $f_2$ are
optimal for these random variables, i.e., there do not exist $f_1'$
(or $f_2'$), such that $\Expt d(X,f_1'(W_1))< \Expt d(X,f_1(W_1))$ (or
$\Expt d(X,f_2'(W_1,W_2,Y))< \Expt d(X,f_2(W_1,W_2,Y))$), because
otherwise we can consider the alternative functions $f_1'$ (or $f_2'$)
without loss of optimality.  Our goal is to show that
$I(X;W_1)+I(X;W_2|YW_1)\geq S^*(D_1,D_2)$, then invoke the rate
distortion theorem, by which the lower bound can be established.

Similar as in \cite{Wynerziv:76}\cite{Kerpez:87}, define the following set
\begin{eqnarray}
A=\{(w_1,w_2):f_2(w_1,w_2,0)= f_2(w_1,w_2,1)\},
\end{eqnarray}
which defines its complement as,
\begin{eqnarray}
A^c=\mathcal{W}_1\times\mathcal{W}_2-A=\{(w_1,w_2):f_2(w_1,w_2,0)\neq f_2(w_1,w_2,1)\}.
\end{eqnarray}

For each $w_1\in \mathcal{W}_1$, define the following two sets
\begin{eqnarray*}
B(w_1)&=&\{w_2\in \mathcal{W}_2:(w_1,w_2)\in A,\, f_1(w_1)=f_2(w_1,w_2,0)\},\\
B^*(w_1)&=&\{w_2\in \mathcal{W}_2:(w_1,w_2)\in A,\, f_1(w_1)\neq f_2(w_1,w_2,0)\}.
\end{eqnarray*}
Notice that for each fixed $w_1^*\in \mathcal{W}_1$, we have
$\mathcal{W}_2=B(w_1^*)\cup B^*(w_1^*)\cup \{w_2:(w_1^*,w_2)\in
A^c\}$, and the three sets are disjoint.  To simplify the notations,
write $P\{(W_1W_2)=(w_1w_2)\}$ as $P_{w_1w_2}$, and $P\{W_1=w_1\}$ as
$P_{w_1}$. Define the following quantity for each $w_1\in
\mathcal{W}_1$
\begin{eqnarray*}
D_{1,w_1}\stackrel{\Delta}{=}\Expt[d(X,\hat{X}_1)|W_1=w_1]=P\{X\neq f_1(w_1)|W_1=w_1\}
\end{eqnarray*}
and define the following quantity for each $(w_1,w_2)\in A$, 
\begin{eqnarray*}
D_{2,w_1w_2}\stackrel{\Delta}{=}\Expt[d(X,\hat{X}_2)|(W_1,W_2)=(w_1,w_2)]=P\{X\neq
f_2(w_1,w_2,0)|(W_1,W_2)=(w_1,w_2)\}.
\end{eqnarray*}
By the Markov string $Y\leftrightarrow X\leftrightarrow (W_1,W_2)$, it follows that for each $w_1\in \mathcal{W}_1$ 
\begin{eqnarray}
H(X|W_1=w_1)=h(D_{1,w_1}),\quad H(Y|W_1=w_1)=h(p*D_{1,w_1}),
\end{eqnarray}
where as before $u*v\stackrel{def}{=}u(1-v)+v(1-u)$.
For each $(w_1,w_2)\in A$, we have
\begin{eqnarray}
H[X|(W_1,W_2=w_1,w_2)]=h(D_{2,w_1w_2}),\quad H[Y|(W_1,W_2)=(w_1,w_2)]=h(p*D_{2,w_1w_2}).
\end{eqnarray}
And furthermore, for each $(w_1,w_2)\in A^c$, we have
\begin{eqnarray}
H[X|(W_1,W_2=w_1,w_2)]&=&h(P\{X\neq f_1(w_1)|W_1=w_1,W_2=w_2\})\nonumber\\ H[Y|(W_1,W_2)=(w_1,w_2)]&=&h(p*P\{X\neq f_1(w_1)|W_1=w_1,W_2=w_2\}).
\end{eqnarray}
We will also need the following quantities
\begin{eqnarray}
\theta\stackrel{\Delta}{=}P\{(W_1,W_2)\in A\}, \quad
\theta_1\stackrel{\Delta}{=}P\{(W_1,W_2)\in \{(w_1,w_2):w_2\in
B(w_1)\}\}.
\end{eqnarray}
Clearly, we have
\begin{eqnarray}
H(X)-H(Y|W_1)&=&1-\sum_{w_1\in\mathcal{W}_1}P_{w_1}H(Y|W_1=w_1)\nonumber\\
&=&1-\sum_{w_1\in\mathcal{W}_1}P_{w_1}h(p*D_{1,w_1})\nonumber\\
&\geq&1-h(p*D_1')
\end{eqnarray}
where we have used the concavity of function $h(p*u)$ in the last step and
\begin{eqnarray*}
D_1'\stackrel{\Delta}{=}\sum_{w_1\in\mathcal{W}_1}P_{w_1}D_{1,w_1}.
\end{eqnarray*}
Furthermore we have
\begin{eqnarray*}
&&H(Y|W_1W_2)-H(X|W_1W_2)\\
&=&\sum_{(w_1,w_2)\in A}P_{w_1,w_2}[H(Y|(W_1,W_2)=(w_1,w_2))-H(X|(W_1,W_2)=(w_1,w_2))]\nonumber\\
&&+\sum_{(w_1,w_2)\in A^c}P_{w_1,w_2}[H(Y|(W_1,W_2)=(w_1,w_2))-H(X|(W_1,W_2)=(w_1,w_2))]
\end{eqnarray*}
The first term can be bounded as follows
\begin{eqnarray}
&&\sum_{(w_1,w_2)\in
A}P_{w_1,w_2}[H(Y|(W_1,W_2)=(w_1,w_2))-H(X|(W_1,W_2)=(w_1,w_2))]\nonumber\\
&=&\sum_{w_1}\sum_{w_2\in
B(w_1)}P_{w_1,w_2}[h(p*D_{2,w_1w_2})-h(D_{2,w_1w_2})]\nonumber\\
&&\qquad\qquad\qquad+\sum_{w_1}\sum_{w_2\in
B^*(w_1)}P_{w_1,w_2}[h(p*D_{2,w_1w_2})-h(D_{2,w_1w_2})]\nonumber\\
&\geq&\theta_1 G(\beta)+(\theta-\theta_1) G(\alpha),
\end{eqnarray}
where as before $G(u)\stackrel{\Delta}{=}h(p*u)-h(u)$, and 
\begin{eqnarray}
\alpha\stackrel{\Delta}{=}\sum_{w_1}\sum_{w_2\in B^*(w_1)}\frac{P_{w_1w_2}}{\theta-\theta_1}D_{2,w_1w_2}, \quad
\beta\stackrel{\Delta}{=}\sum_{w_1}\sum_{w_2\in B(w_1)}\frac{P_{w_1w_2}}{\theta_1}D_{2,w_1w_2},
\end{eqnarray}
and the convexity of function $G(u)$ is used in the last step. Next,
notice the identity that for each $w_1\in\mathcal{W}_1$
\begin{eqnarray}
\label{eqn:identity}
P_{w_1}D_{1,w_1}
&=&P\{X\neq f_1(w_1), W_1=w_1\}\nonumber\\
&=&\sum_{w_2\in B(w_1)} P\{X\neq f_2(w_1,w_2,0), W_1=w_1,W_2=w_2\}\nonumber\\
&&+\sum_{w_2\in B^*(w_1)} P\{X=f_2(w_1,w_2,0), W_1=w_1,W_2=w_2\}\nonumber\\
&&+\sum_{w_2:(w_1,w_2)\in A^c}P\{X\neq f_1(w_1), W_1=w_1,W_2=w_2\}\nonumber\\
&=&\sum_{w_2\in B(w_1)} P_{w_1w_2}D_{2,w_1w_2}+\sum_{w_2\in B^*(w_1)} P_{w_1w_2}(1-D_{2,w_1w_2})\nonumber\\
&&+\sum_{w_2:(w_1,w_2)\in A^c}P_{w_1w_2}P\{X\neq f_1(w_1)|W_1=w_1,W_2=w_2\}.
\end{eqnarray}
It follows that
\begin{eqnarray}
&&\sum_{(w_1,w_2)\in A^c}P_{w_1,w_2}[H(Y|(W_1,W_2)=(w_1,w_2))-H(X|(W_1,W_2)=(w_1,w_2))]\nonumber\\
&=&\sum_{w_1}\sum_{w_2:(w_1,w_2)\in A^c}P_{w_1,w_2}G[P\{X\neq f_1(w_1)|(W_1,W_2)=(w_1,w_2)\}]\nonumber\\
&\geq&(1-\theta) G(\gamma),
\end{eqnarray}
where again the convexity of function $G(u)$ is used, and because of the identity (\ref{eqn:identity}), we have 
\begin{eqnarray}
\gamma&=&\sum_{w_1}\sum_{w_2:(w_1,w_2)\in
A^c}\frac{P_{w_1w_2}}{1-\theta}P\{X\neq
f_1(w_1)|W_1=w_1,W_2=w_2\}\nonumber\\
&=&\frac{D_1'-\theta_1\beta-(\theta-\theta_1)(1-\alpha)}{1-\theta}.
\end{eqnarray}
It was shown in \cite{Kerpez:87}, by a straightforward generalization of the argument in \cite{Wynerziv:76}, that 
\begin{eqnarray}
E[d(X,\hat{X}_2)|(W_1,W_2)\in A^c]\geq p.
\end{eqnarray}
By the hypothesis
\begin{eqnarray*}
&&D_2'\stackrel{\Delta}{=}\theta_1\beta+(\theta-\theta_1)\alpha+(1-\theta)p\leq D_2\\
&&D_1'\leq D_1.
\end{eqnarray*}
Notice that for each $(w_1,w_2)\in A$, $D_{2,w_1w_2}\leq p$, because
otherwise for this $(w_1,w_2)$ pair, making $f_2(w_1,w_2,Y)=Y$ will in
fact reduce the distortion, which contradicts with the optimality of
the decoding function. Thus $0\leq \alpha,\beta\leq p$. Similarly,
$p\leq\gamma\leq 1-p$, because $p\leq P\{X\neq
f_1(w_1)|W_1=w_1,W_2=w_2\}\leq 1-p $, otherwise we can modify the
decoder function $f_2$ to reduce the distortion. Clearly, $0\leq
\theta_1\leq \theta\leq 1$ by definition.

Summarizing the bounds, we have shown that 
\begin{eqnarray}
\label{eqn:leqset}
R_{HB}(D_1,D_2)\geq
\min_{(\alpha,\beta,\theta,\theta_1,D_1')\in\mathcal{Q}_{\geq}}
[1-h(D_1'*p)+(1-\theta)G(\gamma)+\theta_1 G(\beta)+(\theta-\theta_1)
G(\alpha)],
\end{eqnarray}
where the minimization is within the following set
\begin{eqnarray*}
\mathcal{Q}_{\leq}&=&\{(\alpha,\beta,\theta,\theta_1,D_1'):(1-\theta)p\leq
D_1'-(\theta-\theta_1)(1-\alpha)-\theta_1\beta\leq (1-\theta)(1-p),\\
&&0\leq \theta_1\leq \theta\leq 1, \quad 0\leq \alpha,\beta\leq p,
\quad (\theta-\theta_1)\alpha+\theta_1\beta+(1-\theta)p\leq D_2,\quad
D_1'\leq D_1\}.
\end{eqnarray*}
This is not yet the function given in Theorem \ref{theorem:binHB},
because the minimization given there is within the set
\begin{eqnarray*}
\mathcal{Q}_{=}&=&\{(\alpha,\beta,\theta,\theta_1,D_1'):(1-\theta)p\leq
D_1'-(\theta-\theta_1)(1-\alpha)-\theta_1\beta\leq (1-\theta)(1-p),\\
&&0\leq \theta_1\leq \theta\leq 1, \quad 0\leq \alpha,\beta\leq
p,\quad (\theta-\theta_1)\alpha+\theta_1\beta+(1-\theta)p=D_2,\quad
D_1'= D_1\}.
\end{eqnarray*}
This gap will be closed after we give the forward test channel structure. \hfill$\square$

\noindent{\em The upper bound}

\begin{table}[tb]
\centering
\begin{tabular}{|c|c|c|c|c|}\hline
 &\multicolumn{2}{c|}{$w_1=0$}&\multicolumn{2}{c|}{$w_1=1$}\\\cline{2-5}
 &$x=0$&$x=1$&$x=0$&$x=1$\\\hline $w_2=0$&
 $0.5\theta_1(1-\beta)$&$0.5\theta_1\beta$&$0.5(\theta-\theta_1)(1-\alpha)$&$0.5(\theta-\theta_1)\alpha$\\\hline
 $w_2=1$&$0.5(\theta-\theta_1)\alpha$&$0.5(\theta-\theta_1)(1-\alpha)$&$0.5\theta_1\beta$&$0.5\theta_1(1-\beta)$\\\hline
 $w_2=2$&$0.5(1-\theta)(1-\gamma)$&$0.5(1-\theta)\gamma$&$0.5(1-\theta)\gamma$&$0.5(1-\theta)(1-\gamma)$\\\hline\hline
 $p(x,w_1)$&$0.5(1-D_1)$&$0.5D_1$&$0.5D_1$&$0.5(1-D_1)$\\\hline
\end{tabular}
\caption{Joint distribution $p(x,w_1,w_2)$ and the marginal $p(x,w_1)$.\label{table:pmf}}
\end{table}

We explicitly construct the random variables with joint pmf given in
Table \ref{table:pmf}. It is straightforward to verify that it is a
valid pmf, given the conditions in the definition of
$S_{D_1}(\alpha,\beta,\theta,\theta_1)$. Furthermore, the rate
$I(X;W_1)+I(X;W_2|W_1Y)$ is exactly
$S_{D_1}(\alpha,\beta,\theta,\theta_1)$. The decoding functions are
$f_1(W_1)=W_1$ and $f_2(W_1,W_2,Y)=W_2$ if $W_2\neq 2$, otherwise
$f_2(W_1,W_2,Y)=Y$. This establishes the upper bound.

Now we show that the gap aforementioned in the proof of the lower
bound can be closed. Suppose that the parameters that minimize the
right hand side of (\ref{eqn:leqset}) are
$(\alpha,\beta,\theta,\theta_1,D_1')$, and furthermore $D_1'<D_1$. The
set of random variables $W_1',W_2'$ can be constructed as given in Table
\ref{table:pmf} with $D_1'$ replacing $D_1$. By the lower bound
established above, we have
\begin{eqnarray}
R_{HB}(D_1,D_2)\geq I(X;W_1')+(X;W_2'|W_1'Y).
\end{eqnarray}
Consider a random variable $W_1''=W_1'\oplus N$, where $N$ is a
Bernoulli random variable independent of everything else with
$P(N=1)=\eta$ such that $\eta*D_1'=D_1=D_1''$, which is valid since $\max\{D_1,D_1'\}\leq \frac{1}{2}$. 
Let $W_2''=(W_1',W_2')$, and we
have $(W_1'',W_2'')\in P(D_1,D_2)$. Clearly, $W_1''\leftrightarrow
W_1'\leftrightarrow X\leftrightarrow Y$, and $W_1''\leftrightarrow
W_1'\leftrightarrow W_2'$. Thus by the rate distortion theorem for this
problem
\begin{eqnarray}
I(X;W_1'')+I(X;W_2''|W_1''Y)\geq R_{HB}(D_1,D_2).
\end{eqnarray}
Notice that
\begin{eqnarray*}
&&I(X;W_1')+I(X;W_2'|W_1'Y)\nonumber\\
&\stackrel{(a)}{=}& I(X;W_1',W_1'')+I(X;W_1'',W_2'|W_1'Y)\\
&=&I(X;W_1'')+I(X;W_1'|W_1'')+I(X;W_2'|W_1'W_1''Y)\\
&\stackrel{(b)}{=}&I(X;W_1'')+I(X;W_1'|W_1'')+I(X;W_1'W_2'|W_1''Y)-I(X;W_1'|W_1''Y)\\
&\stackrel{(c)}{=}&I(X;W_1'')+I(X;W_1'W_2'|W_1''Y)+I(Y;W_1'|W_1'')\\
&=&I(X;W_1'')+I(X;W_1'W_2'|W_1''Y)+h(p*D_1'')-h(p*D_1')\\
&>&I(X;W_1'')+I(X;W_1'W_2'|W_1''Y)
\end{eqnarray*}
where $(a)$ and $(c)$ follow because of the Markov chain $W_1''\leftrightarrow W_1'\leftrightarrow X \leftrightarrow Y$, $(b)$ is by applying chain rule to the last term in the previous line, and the last step is because $p<0.5$ and $D_1'<D_1=D_1''\leq 0.5$. However, this implies
\begin{eqnarray*}
&&I(X;W_1'')+I(X;W_1'W_2'|W_1''Y)\geq R_{HB}(D_1,D_2)\nonumber\\
&&\qquad\qquad\qquad\qquad\qquad\geq I(X;W_1')+(X;W_2'|W_1'Y)>I(X;W_1'')+I(X;W_1'W_2'|W_1''Y)
\end{eqnarray*}
which is a contradiction. Thus we conclude that the minimum must be achieved with $D_1'=D_1$. 

Next we show that the constraint
$(\theta-\theta_1)\alpha+\theta_1\beta+(1-\theta)p\leq D_2$ can be met
with equality without loss of optimality; i.e.,
\begin{eqnarray}
&&\min_{(\alpha,\beta,\theta,\theta_1,D_1')\in\mathcal{Q}_{\geq}}
[1-h(D_1'*p)+(1-\theta)G(\gamma)+\theta_1 G(\beta)+(\theta-\theta_1)
G(\alpha)]\nonumber\\
&=&\min_{(\alpha,\beta,\theta,\theta_1,D_1')\in\mathcal{Q}_{=}}
[1-h(D_1'*p)+(1-\theta)G(\gamma)+\theta_1 G(\beta)+(\theta-\theta_1)
G(\alpha)].
\end{eqnarray}
Suppose otherwise, such that the parameters
$(\alpha,\beta,\theta,\theta_1,D_1)$ minimizing the right hand side of
Eqn. (\ref{eqn:leqset}) satisfy
$(\theta-\theta_1)\alpha+\theta_1\beta+(1-\theta)p< D_2$, and any 
parameters $(\alpha,\beta,\theta,\theta_1,D_1)\in\mathcal{Q}_{=}$ will
result in a strict increase in the rate.  If $\theta=0$, the
contradiction is trivial: either $\alpha$ or $\beta$ can be increase
to reduce the rate. When $\theta<1$, but $\alpha,\beta<p$,
$\gamma\in(p,0.5)\cup(0.5,1-p)$ and $0<\theta_1<\theta$, it is also
trivial to construct such parameters, by disturbing (incrementally)
$\alpha$ or $\beta$.  Thus the only remaining cases are the follows,
and we will ignore the term $1-h(p*D_1)$ in the sequel:
\begin{itemize}
\item $p\leq \gamma\leq0.5$, $\alpha=p$ and $\theta_1<\theta$. In this case, notice that 
\begin{eqnarray*}
(1-\theta)G(\gamma)+\theta_1 G(\beta)+(\theta-\theta_1)G(\alpha)
&=&(1-\theta)G(\gamma)+\theta_1 G(\beta)+(\theta-\theta_1)G(1-\alpha)\\
&>&(1-\theta_1)G(\frac{D_1-\theta_1\beta}{1-\theta_1})+\theta_1 G(\beta),
\end{eqnarray*} 
where the inequality is due to the strict convexity of
$G(u)$. Furthermore, notice that
$p\leq\frac{D_1-\theta_1\beta}{1-\theta_1}\leq 1-p$, since it is a
convex combination of $\gamma$ and $1-p$.  However, this implies the
set of parameters $(p,\beta,\theta_1,\theta_1)$ strictly improves over
the minimum, which is a contradiction.
\item $p\leq \gamma\leq0.5$ and $\theta=\theta_1$. Let $\epsilon$ be a
small positive quantity to be specified later. First notice the
condition implies that $\beta<p$ for any $D_2<p$, then
\begin{eqnarray*}
(1-\theta)G(\gamma)+\theta G(\beta)
&=&(1-\theta-\epsilon)G(\gamma)+\epsilon G(\gamma)+\theta G(\beta)\\
&>&(1-\theta-\epsilon)G(\gamma)+(\theta+\epsilon) G(\beta'),
\end{eqnarray*}
where the inequality is due to the strictly convexity of $G(u)$ and
\begin{eqnarray}
\beta'\stackrel{\Delta}{=}
\frac{\epsilon(D_1-\theta\beta)}{(\epsilon+\theta)(1-\theta)}+\frac{\theta\beta}{\epsilon+\theta}.
\end{eqnarray}
Notice further that 
\begin{eqnarray}
\gamma=\frac{D_1-\theta\beta}{1-\theta}=\frac{D_1-(\theta+\epsilon)\beta'}{1-\theta-\epsilon}
\end{eqnarray}
thus by choosing a sufficient small $\epsilon>0$, the following two conditions can be satisfied simultaneously,
\begin{eqnarray}
(\theta+\epsilon)\beta'+(1-\theta-\epsilon)p=\theta\beta+(1-\theta-\epsilon)p+\epsilon(\gamma-p)\leq D_2, \quad
\beta'\leq p.
\end{eqnarray}
This implies that $(p,\beta',\theta+\epsilon,\theta+\epsilon)$
strictly improves over the minimum, which is a contradiction.
\item $0.5\leq \gamma\leq 1-p$, $\beta=p$ and $\theta_1>0$. The
contradiction is similarly constructed as the first case.
\item $0.5\leq \gamma\leq 1-p$ and $\theta_1=0$. This is an impossible case, since $\alpha\leq p$ and $D_1\leq 0.5$.  
\item $\lambda=0.5$ and $0<\theta_1<\theta$, $0\leq\alpha,\beta<p$. In
this case, perturbing $\alpha,\beta$ together incrementally gives a
contradiction.
\end{itemize}
Thus there is no loss of optimality by replacing the optimization set
$\mathcal{Q}_{\leq}$ with $\mathcal{Q}_{=}$, and this completes the
proof.

\hfill$\square$

\subsection{Proof of Corollary \ref{corollary:bin}}

Notice that for any $(\alpha,\beta,\theta,\theta_1)$, 
\begin{eqnarray*}
S_{D_1}(\alpha,\beta,\theta,\theta_1)
&\geq&1-h(D_1*p)+(\theta-\theta_1)G(\alpha)+\theta_1G(\beta)\\
&\geq&1-h(D_1*p)+\theta G(\beta')
\end{eqnarray*}
where
$\beta'\stackrel{\Delta}{=}\frac{(\theta-\theta_1)\alpha+\theta_1\beta}{\theta}$,
and the first inequality is due to the non-negativity of function
$G(u)$, while the second inequality is due to its convexity.
Furthermore, the constraint is satisfied with
\begin{eqnarray*}
D_2=(\theta-\theta_1)\alpha+\theta_1\beta+(1-\theta)p=\theta\beta'+(1-\theta)p.
\end{eqnarray*}
Let $(\alpha,\beta,\theta,\theta_1)$ be the set of parameters achieving the minimum. 
Then by Theorem \ref{theorem:binHB}, we have
\begin{eqnarray*}
R_{HB}(D_1,D_2)= S_{D_1}(\alpha,\beta,\theta,\theta_1)\geq [1-h(D_1*p)+\theta G(\beta')],
\end{eqnarray*}
where $D_2=\theta\beta'+(1-\theta)p$. Moreover $0\leq\beta'\leq p$,
because both $\alpha$ and $\beta$ are in this range, and $\beta'$ is
the convex combination of them. Thus
\begin{eqnarray*}
R_{HB}(D_1,D_2)\geq 1-h(D_1*p)+\min_{D_2=\theta\beta'+(1-\theta)p}[\theta G(\beta')],
\end{eqnarray*}
with the minimization range $0\leq \beta'\leq p$ and $0\leq \theta\leq
1$. Comparing it with the rate distortion function $R^*_{X|Y}(D)$ of
(\ref{eqn:wzBin}) establishes the claim.  \hfill$\square$

\subsection{Proof of Corollary \ref{corollary:special}}

In \cite{Wynerziv:76}, it was proved that when $D_2\leq d_c$,
$R^*_{X|Y}(D_2)=G(D_2)$, and by Corollary \ref{corollary:bin},
$R_{HB}(D_1,D_2)\geq 1-h(D_1*p)+G(D_2)$ for this case. To show
$R_{HB}(D_1,D_2)\leq 1-h(D_1*p)+G(D_2)$, consider the following test
channel. Let $W_2$ be the output of a binary symmetric channel (BSC)
with crossover probability $D_2$ and input $X$, let $W_1$ be the
(cascade) output of a BSC with crossover probability $\eta$ with input
$W_2$, such that $\eta*D_2=D_1$; such an $\eta$ always exists because
$D_2\leq D_1$. It can then be easily verified that
\begin{eqnarray}
I(X;W_1)+I(X;W_2|W_1,Y)=1-h(D_1*p)+G(D_2)
\end{eqnarray}
and the distortion is $D_1$ and $D_2$ by taking $f_1(W_1)=W_1$ and
$f_2(W_1,W_2,Y)=W_2$. The rate distortion theorem for this problem implies
that $R_{HB}(D_1,D_2)\leq 1-h(D_1*p)+G(D_2)$, which completes the
proof. \hfill$\square$

\bibliographystyle{ieeetr}
\bibliography{UMDSQ}
\end{document}